\documentclass[aps,twocolumn,prl,showpacs,showkeys,preprintnumbers,superscriptaddress,nobibnotes,floatfix,nofootinbib]{revtex4-2}

\usepackage{orcidlink}
\usepackage{amsmath}
\usepackage{amsfonts}
\usepackage{amssymb}
\usepackage{mathrsfs}
\usepackage{bm} %  \bm{\alpha}
\usepackage{bbm} %  \mathbbm{1}
\usepackage{slashed} % \slashed 
\usepackage{physics} %  \grad, \curl
\usepackage{esvect} %  \vec{v}
\usepackage[version=4]{mhchem}
\usepackage{xcolor} %  color
\usepackage{graphicx} % 
\usepackage{array} %  graphicx 
\usepackage{url} % 
\usepackage{amsthm} % / \begin{theorem}
\usepackage{enumerate} %  \begin{enumerate}[(1)]
\usepackage{enumitem} % 
\usepackage{booktabs} %  \toprule, \midrule
\usepackage{siunitx} %  \SI{10}{\meter}
\usepackage{mdframed} % 
\usepackage{tcolorbox} % 
\usepackage{fontawesome5} %  \faFilePdf
\usepackage{braket} %  \braket{\psi|\phi}
\usepackage{setspace} % 
\usepackage{multirow} % 
\usepackage{upgreek} % upgreek  \upalpha
\usepackage{xspace} % 
\usepackage[normalem]{ulem} % For \sout
\usepackage{array}
\usepackage{natbib} % revtex4-1 
\usepackage{hyperref} % 
\hypersetup{
    colorlinks=true,
    linkcolor=blue,
    citecolor=blue,
    filecolor=magenta,
    urlcolor=blue,
    breaklinks=true
}
\usepackage[all]{hypcap} % /
\usepackage[capitalise]{cleveref} % 

\setlist[description]{leftmargin=\parindent,labelindent=\parindent} % 

\newcommand{\Op}[1]{\hat{#1}}
\newcommand{\D}{\Op{D}}
\newcommand{\R}{\Op{R}}

\newcommand{\adag}{\Op{a}^\dagger}
\newcommand{\aop}{\Op{a}}
\newcommand{\Impart}{\text{Im}}

\newcommand{\sinc}{\text{sinc}}
\newcommand{\prlsection}[2]{{\it\textbf{#1}{#2}}---}
\definecolor{gesfred}{rgb}{1,0,0}

\definecolor{gesfblue}{rgb}{0.08,0.42,0.76}

\newcounter{appsec}
\renewcommand{\theappsec}{\Alph{appsec}}
\newcommand{\appsection}[2]{%
    \refstepcounter{appsec}%
    \section{Appendix \theappsec\ -- #2}%
    \phantomsection%
    \label{#1}%
}
\newcommand{\appsectionnolabel}[1]{%
    \refstepcounter{appsec}%
    \section{Appendix \theappsec\ -- #1}%
}

\graphicspath{{figs/}}

\begin{document}
\preprint{KEK-QUP-2026-0006, IPMU26-0012}
%\preprint{KEK-QUP-2026-0006}
% =====================================================================
% TITLE & AUTHORS
% =====================================================================
\title{Enhanced Dark Matter Quantum Sensing via Phase-Space Geometric Interferometry}

\author{Xiaolin Ma~\orcidlink{0009-0007-1994-9493}} 
\altaffiliation{\href{mailto:xlmphy@post.kek.jp}{xlmphy@post.kek.jp, corresponding}} 
\affiliation{International Center for Quantum-field Measurement Systems for Studies of the Universe and Particles (QUP, WPI),
High Energy Accelerator Research Organization (KEK), Oho 1-1, Tsukuba, Ibaraki 305-0801, Japan}

\author{Jie Sheng}
\altaffiliation{\href{mailto:jie.sheng@ipmu.jp}{jie.sheng@ipmu.jp, corresponding}} 
\affiliation{
Kavli IPMU (WPI), UTIAS, University of Tokyo, Kashiwa, 277-8583, Japan}

\begin{abstract}
We propose a quantum sensing protocol for coupled qubit-oscillator systems that enhances the sensitivity by exploiting a geometric phase for dark matter searches. Instead of letting the cavity evolve freely under a weak dark matter background, we combine large coherent displacements and squeezing operations within the evolution protocol, thereby mapping the signal onto an enhanced quantum phase determined by the geometric area enclosed in phase space.
Relative to the conventional free-evolution scheme, the protocol enhances the quantum Fisher information and improves the projected sensitivities to cavity-based detection of dark-photon and axion dark matter.

\end{abstract}
\maketitle

\prlsection{Introduction}{.}
The nature of dark matter (DM) remains one of the most profound unsolved mysteries in fundamental physics~\cite{Planck:2018vyg,Cirelli:2024ssz}. While its gravitational influence is well-established, its particle identity is unknown~\cite{Bertone:2004pz,Bertone:2016nfn}. Among the most compelling candidates are weakly interacting slim particles (WISPs)~\cite{Arza:2026rsl}, such as axions~\cite{PhysRevD.16.1791,PhysRevLett.38.1440,PhysRevLett.40.279,PhysRevLett.40.223} and dark photons~\cite{Holdom:1985ag,Arias:2012az}, which are theorized to couple feebly to Standard Model particles.
The search for these elusive particles has driven the development of exquisitely sensitive detectors~\cite{PhysRevLett.51.1415,PhysRevD.32.2988,PhysRevLett.112.131301,PhysRevX.4.021030,PhysRevD.89.043522,Chaudhuri:2014dla,Graham:2017ivz,Berlin:2019ahk,Chen:2022quj,PhysRevD.110.115021,Chen:2024aya,Fierlinger:2024aik,Fierlinger:2024rdj,Green:2025qgs,Wei:2023rzs,Xu:2023vfn,Huang:2026tgv}, effectively turning the hunt for new physics into a challenge at the frontiers of quantum metrology~\cite{Degen:2016pxo}.

Resonant microwave cavities, or haloscopes, are cornerstones of this effort, designed to detect the conversion of DM particles into faint electromagnetic signals~\cite{PhysRevLett.51.1415,PhysRevD.32.2988,RevModPhys.75.777,Jaeckel:2007ch,RevModPhys.93.015004,Arias:2012az,Caputo:2021eaa}. In conventional haloscopes, 
the DM field deposits a tiny amount of energy into the cavity, producing a small displacement of the cavity mode from the vacuum state.
These experiments typically probe frequencies in the GHz regime~\cite{Aybas:2026rwu}, corresponding to DM masses in the $\mu$eV range, which is well motivated for both axion~\cite{Preskill:1982cy,Abbott:1982af,Dine:1982ah,Sheng:2025sou,Georis:2025kzv} and dark photon DM~\cite{Nelson:2011sf,Kolb:2020fwh,Nakai:2020cfw}.
However, their sensitivity is ultimately limited by the Standard Quantum Limit (SQL)~\cite{Clerk:2008tlb,Degen:2016pxo},
set by the vacuum fluctuations of the cavity mode.
This has motivated extensive studies of quantum-control strategies for improving weak-signal detection~\cite{Chen:2023swh,Giovannetti:2004cas,Zheng:2025qgv,Ma:2025orp,Suri:2025hqm,Wang:2026yhs}.

Geometric-phase sensing in coupled qubit--oscillator systems for enhancing the sensitivity of quantum sensing was developed in ~\cite{Suri:2025hqm}.
In this Letter, building on this concept, we introduce a cavity-based DM sensing protocol in which a Ramsey-echo sequence maps the weak DM-induced cavity displacement onto a relative qubit phase. Coherent pumping and squeezing enlarge the controlled cavity displacement, which increases the enclosed phase-space area and hence the measured geometric phase. We derive the resulting QFI, its trade-off with cavity-loss-induced dephasing, and the projected dark photon and axion sensitivities obtained from an explicit analysis.

\prlsection{System Hamiltonian with Dark Matter Coupling}{.} 
We consider the dispersively coupled qubit–oscillator system implemented in circuit quantum electrodynamics (cQED) architectures~\cite{PhysRevA.69.062320,Clerk2020Hybrid,Krantz2019QuantumEngineerGuide}, where high-fidelity qubit readout is routinely available~\cite{PhysRevApplied.17.044016,Kim:2025ywx}. The oscillator is a high-Q microwave cavity, and the qubit is a superconducting transmon, whose core element is the Josephson junction, a device widely employed as a quantum sensor~\cite{Chen:2022quj,PhysRevD.110.115021,Chen:2023swh,Chen:2024aya,Cheng:2024yrn,Cheng:2024zde,Chao:2024owf,Gao:2025ryi,Sheng:2025ygn}. 
An ultralight-DM background induces an effective electric current $\vec J_{\rm eff}$ that couples to the cavity through $V(t)=\int_V \vec{A}_{\rm EM}\cdot \vec J_{\rm eff}\,dV$. After decomposing the cavity field into modes and transforming to the rotating frame, the interaction Hamiltonian is~\footnote{Here we choose not to put the dispersive term into the unperturbed term in $H_0$ as the ``dispersive dressed" interaction picture~\cite{Suri:2025hqm}, and instead choose the hierarchy perturbation picture is to make the operation like displacement and squeezing more clear.}
\begin{equation}
    H_I(t) \equiv H_0 + V(t) = \frac{\chi}{2} \Op{\sigma}_z \adag \aop + \left[ A \adag e^{i(\Delta t - \phi_1)} + \text{h.c.} \right].
\label{hamiltonian}
\end{equation}
The first term $H_0 \equiv \Omega_s \adag \aop$ with $\Omega_s \equiv s\chi/2$ represents the qubit-conditioned cavity rotation.
Here, $\adag$ and $\aop$ denote the creation and annihilation operators, respectively, for the $\mathrm{TM}_{010}$ cavity mode, and $\chi$ is the dispersive coupling rate.
The DM--cavity coupling amplitude $A$ is defined as $A=A_D$ for dark-photon DM and $A=A_a$ for axion DM, as given in Eqs.~\eqref{couplingA} and \eqref{couplinga}, respectively.
Throughout this work, we use the convention $\Op{\sigma}_z\ket{e}=+\ket{e}$ and $\Op{\sigma}_z\ket{g}=-\ket{g}$, so that $s=+1$ labels the excited-state branch and $s=-1$ labels the ground-state branch, analogous to an effective spin state.

The driving term $V(t)$ represents the DM-cavity coupling with detuning $\Delta \equiv \omega_c - \omega_D$ and random phase $\phi_1$. For dark-photon DM with kinetic mixing parameter $\epsilon$ relative to the Standard Model photon, the drive amplitude is~\cite{Arias:2012az}
\begin{equation}
A_D = \epsilon m_{A'} \sqrt{\rho_{\mathrm{DM}} V_{\mathrm{eff}} / \omega_c},
\label{couplingA}
\end{equation}
where $\rho_{\text{DM}} \simeq 0.45\,\text{GeV}/\text{cm}^3$ is the local DM density and $V_{\mathrm{eff}} \equiv | \int_V d^3x \bm{\epsilon}(\bm{x})\cdot \bm{\epsilon}_D e^{-i\bm{k}\cdot\bm{x}} |^{2}$ is the mode overlap volume~\cite{Arias:2012az}. 
For axion DM coupling to the cavity electric field via a magnetic field $B_0$ with coupling strength $g_{a\gamma\gamma}$, the amplitude is~\cite{PhysRevLett.51.1415,RevModPhys.75.777}
\begin{equation}
A_a = g_{a\gamma\gamma} B_0 \sqrt{\rho_\text{DM} \omega_c C V / m_a^2},
\label{couplinga}
\end{equation}
where $C \sim \mathcal{O}(1)$ is the form factor~\cite{RevModPhys.75.777}. Derivations are detailed in Appendix~\ref{app:DM_coupling}.

The evolution under the DM drive can be derived using the Magnus expansion in the interaction picture. The first-order term in $A$ generates the effective displacement operator $\hat D$, whereas the second-order term generates the state-dependent phase $e^{i\Phi_G(s)}$. Details are given in Appendix~\ref{app:EvolutionOp}. For an interval of duration $\tau=t_f-t_i$, the resulting evolution operator is
\begin{align}
    U(t_i,t_f)= e^{i \Phi_G(s)} \R(\Omega_s \tau) \D(\delta_s).
\label{evo_op}
\end{align}
Here, $\R(\Omega_s\tau)\equiv\exp(-i\Omega_s\tau\adag\aop)$.
The effective drift $\delta_s$ is intrinsically linked to the DM field:
\begin{align}
    \delta_s(t_f,t_i)\equiv -\frac{A e^{-i\phi_1}}{\Omega_{\rm eff,s}} e^{i\Omega_{\rm eff,s}\tau} \left[ 1 - e^{-i\Omega_{\rm eff,s}\tau} \right] e^{i\Delta t_i},
    \label{delta_s}
\end{align}
The overall detune frequency $
\Omega_{\mathrm{eff},s}
\equiv \Delta+\Omega_s
=\Delta+s\chi/2$ takes into account of the frequency shift from the cavity-qubit dispersive coupling.
Acting on the vacuum, the displacement operator produces the coherent state $\hat D(\delta_s)|0\rangle=|\delta_s\rangle$.
The phase is 
\begin{align}
    \Phi_G(s)=- \frac{A^2}{\Omega_{\text{eff}, s}} \left[ \tau - \frac{\sin(\Omega_{\text{eff}, s} \tau)}{\Omega_{\text{eff}, s}} \right]. 
\end{align}
One can see the total evolution operator naturally decomposes into a spin-dependent phase $\Phi_G(s)$ of order $A^2$, a dispersive dynamical rotation $\R(\Omega_s\tau)$, and a DM-induced coherent displacement $\D(\delta_s)$.

In the laboratory frame, the cavity mode initialized in the ground state $|\psi(0)\rangle=\ket{0}$ would evolve to the final state 
$\ket{\Psi_i(\tau)}\approx \ket{0}+ c_1(\tau)\ket{1}$ 
with 
\begin{equation}
    c_1 (\tau)\approx -iAe^{i \Phi_G(s)} e^{-i\big[\omega_c+\omega_D+\Omega_s\big]\tau/2-i\phi_1}\frac{\sin(\Omega_{\rm eff,s}\tau/2)}{\Omega_{\rm eff,s}/2}
\end{equation}
in the weak-drive limit, for which $|\delta_s\rangle\simeq|0\rangle+\delta_s|1\rangle$.
After averaging over the DM ensemble, the probability of finding the cavity in $|1\rangle$ is $P(\tau)\equiv\langle |c_1|^2\rangle$, with
\begin{equation}
    P(\tau) = \int d\omega ~f_{\rm DM}(\omega) |A|^2 \left(\frac{\sin(\Omega_{\rm eff,s}\tau/2)}{\Omega_{\rm eff,s}/2}\right)^2.
\end{equation}
If only one specific spin state of the qubit is considered, the $\Omega_{\rm eff,s}$ becomes the effective detune between DM and the cavity frequency.
This is the typical observable in the usual cavity based DM detection experiments~\cite{Zheng:2025qgv,SHANHE:2023kxz}. For a cavity with a quality factor $Q_c$ exceeding the DM quality factor $Q_{\text{DM}} \sim 10^6$, the signal power no longer increases with $Q_c$ and is limited by the DM linewidth.

\prlsection{Geometric Protocol Sequence and Signal Phase}{.}
Instead of letting the system freely evolve in the DM background,
our protocol uses squeezing and a spin-echo sequence to amplify the weak DM signal. It consists of three sequential blocks that steer the oscillator around an effective phase-space loop, as illustrated in Fig.~\ref{fig:protocol}. The detailed derivation is given in Appendix~\ref{app:protocol}.
Throughout, we assume that the durations of the squeezing, displacement, and qubit pulses are negligible compared with $\tau_0$. Qubit rotations and cavity displacements with nanosecond-scale microwave pulses have been demonstrated in cQED systems~\cite{Chou:2018teleportation}, while intracavity squeezing dynamics have been realized on microsecond scales~\cite{Cai:2025eve}. These control scales are short compared with the benchmark free-evolution time $\tau_0\sim0.5\,\mathrm{ms}$ at $\omega_c/2\pi\sim1\,\mathrm{GHz}$.

\begin{figure*}[htp!]
    \centering
    \includegraphics[width=0.99\linewidth]{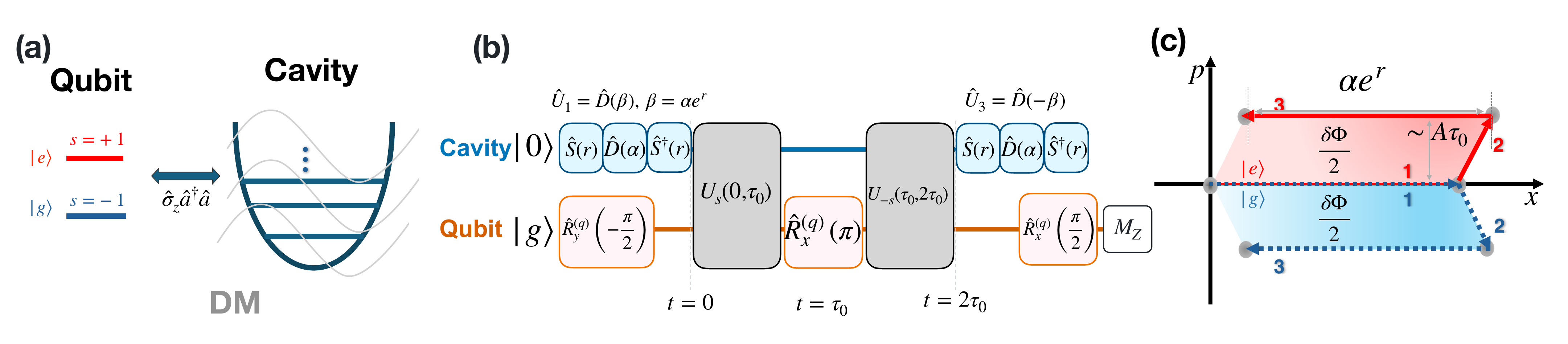}
    \caption{Geometric-phase sensing sequence. (a) A transmon qubit is dispersively coupled to a cavity mode that is weakly driven by the DM background, the qubit state $s=\pm1$ shifts the cavity evolution through the interaction $\chi\Op{\sigma}_z\adag\aop/2$. (b)  Cavity and qubit operations of protocol. At $t=0$, the qubit rotation $R_y^{(q)}(-\pi/2)$ prepares the Ramsey superposition while the cavity operation $U_1=\hat S^\dagger(r)\hat D(\alpha)\hat S(r)=\hat D(\beta)$ generates the controlled displacement $\beta=\alpha e^r$. The joint system then evolves for $\tau_0$ on branch $s$, the qubit echo $R_x^{(q)}(\pi)$ switches the conditional label to $-s$, and a second interval of duration $\tau_0$ follows. At $t=2\tau_0$, the cavity operation $U_3=\hat D(-\beta)$ reverses the controlled displacement, after which $R_x^{(q)}(\pi/2)$ and the population measurement $M_Z$ read out the Ramsey phase. (c) Conditional phase-space trajectories for $s=+1$ (red) and $s=-1$ (blue). The numbered segments correspond to the three blocks in panel (b), and their branch-differential oriented area gives $\delta\Phi$.}
    \label{fig:protocol}
\end{figure*}    

We denote qubit rotations by $R_\mu^{(q)}(\theta)\equiv\exp(-i\theta\Op{\sigma}_\mu/2)$. At $t=0$, $R_y^{(q)}(-\pi/2)$ prepares the qubit state $(\ket{g}+\ket{e})/\sqrt{2}$, while Block~1 acts on the cavity. A strong cavity pump produces the displacement $\alpha$, and conjugation by the cavity squeezing operations $\hat S(r)$ amplifies the effective displacement. The unitary evolution can be written as
\begin{align}
    U_1=\hat{S}^\dagger(r)\hat D(\alpha) \hat S(r)=\hat D(\alpha e^r)\equiv \hat{D}(\beta).
\end{align}
For real $\alpha$, the squeezing operation enhances the controlled displacement to $\beta\equiv\alpha e^r$.
A direct displacement $\hat D(\beta)$ would give the same sensing performance; the squeezing sequence is one implementation route for producing a large effective displacement. We also note that the squeezing operations in this protocol are not used to prepare a squeezed state.

In the second block ($U_2$), we let the coupled cavity-spin system evolve freely over the interval $t \in [0, 2\tau_0]$. During the first $\tau_0$, the dynamics are conditioned on the qubit eigenstate $s$. Over this period, the DM weakly drives the cavity, generating a small displacement $\delta_{1a}$. A spin-echo $\pi$ pulse then inverts the qubit branch, and a second free-evolution interval of duration $\tau_0$ conditioned on $-s$ generates the drift $\delta_{1b}$.
At $t=\tau_0$, the qubit echo pulse $R_x^{(q)}(\pi)$ reverses the dispersive sign because $R_x^{(q)}(\pi)\Op{\sigma}_zR_x^{(q)\dagger}(\pi)=-\Op{\sigma}_z$. The full joint sequence for Block~2 is $U(\tau_0,2\tau_0)R_x^{(q)}(\pi)U(0,\tau_0)$. In the branch-resolved notation below, the physical pulse is represented by $s\to-s$.
The branch-resolved evolution operator is
\begin{align}
    U_2=U_{-s}(\tau_0,2\tau_0)U_s(0,\tau_0)
\end{align}

An imperfect reverse displacement leaves a residual cavity amplitude and can add a qubit-correlated Ramsey phase. We state the corresponding calibration inequalities after Eq.~\eqref{eta}; their derivation is given in Appendix~\ref{app:reversal}.

At $t=2\tau_0$, Block~3 applies the inverse displacement operation to the cavity,
\begin{align}
    U_3=\hat{S}^\dagger(r) \hat D(-\alpha) \hat S(r)=\hat D^\dagger(\beta),
\end{align}
which reverses the controlled displacement of Block~1 and completes the effective geometric sequence. The qubit analysis pulse $R_x^{(q)}(\pi/2)$ then maps the accumulated Ramsey phase onto the population measured by $M_Z$.

The total evolution operator is obtained by multiplying the above blocks in sequence. By repeatedly using the commutation relations between displacement operators 
\begin{equation}
    D(a)D(b)
=
e^{\,i\,\mathrm{Im}(a^*b)}\,
D(a+b)
\label{DDoperation}
\end{equation}
we can simplify the product and extract an overall phase evolution together with the net drift.
Keeping only the spin-dependent phase, it can be written as:
\begin{align}
    U_{\rm tot}=U_3U_2U_1=\hat{D}(\Sigma)e^{is\delta \Phi/2}.
\end{align}
Although $\hat U_3$ is qubit independent and therefore does not generate the Ramsey phase by itself, it plays an important role to cancel the displacement operations between Blocks~1 and 3 and leave the residual displacement $\Sigma\approx A\tau_0$ generated only by the DM interaction. Thus Block~3 suppresses the large coherent cavity background associated with the controlled displacement as shown later.
The two qubit branches also acquire the relative phase
\begin{align}
    \delta\Phi &= A \beta \chi \tau_0^2\sin(\Delta \tau_0-\phi_1)\nonumber\\ &\times\sinc\left(\frac{(\Delta+\chi/2)\tau_0}{2}\right) \sinc\left(\frac{(\Delta-\chi/2)\tau_0}{2}\right).
\label{dPhi}
\end{align}
The full operator decomposition confirms the phase separation. The two equal free-evolution intervals, connected by $R_x^{(q)}(\pi)$, cancel the net dispersive rotation. The single-interval Magnus terms enter as $\Phi_G(s)+\Phi_G(-s)$, which is common to the two Ramsey branches and therefore cancels from their relative phase. To first order in the DM drive, the remaining branch-dependent phase comes from the noncommutative product $\hat D(-\beta)\hat D(\Sigma_s)\hat D(\beta)$; for real $\beta$, the phase on branch $s$ is $2\beta\,\operatorname{Im}\Sigma_s$, and the branch difference is precisely Eq.~\eqref{dPhi}. 

\begin{figure}[t]
    \includegraphics[width=0.99\linewidth]{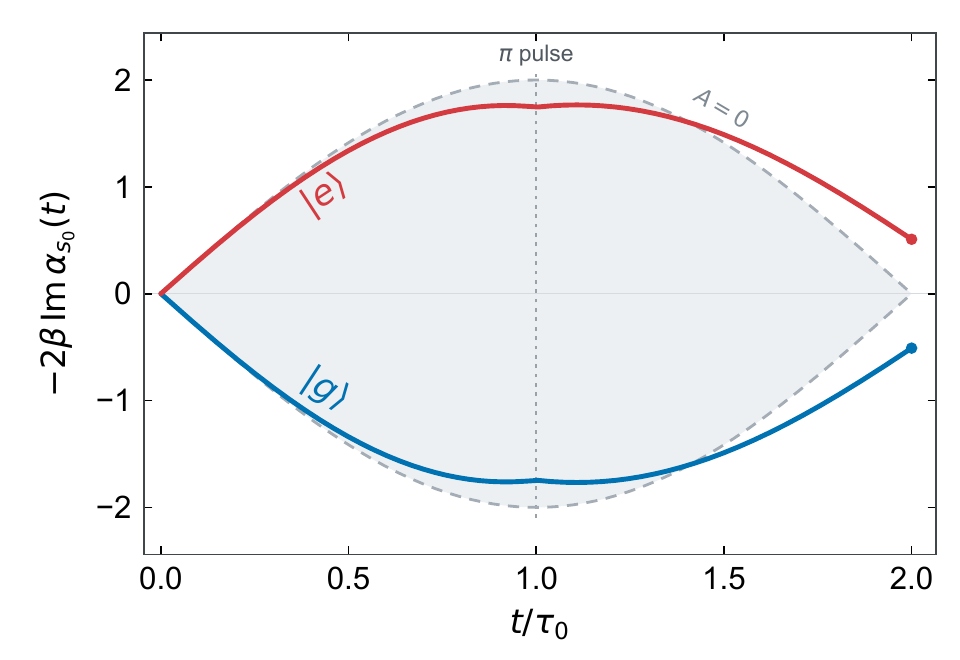}
    \caption{Time--quadrature trajectories for $\beta=1$, $\chi\tau_0=\pi$, $A\tau_0=0.2$, $\Delta=0$, and $\phi_1=\pi/2$. Blue and red curves track the qubit branches initially in $|g\rangle$ and $|e\rangle$, respectively, with a $\pi$ pulse at $t=\tau_0$. Gray dashed curves show the undriven ($A=0$) reference trajectories. The shaded region represents their separation integrated over the normalized time $t/\tau_0$; multiplying this reference area by the drive amplitude $A$ gives the linear DM-induced phase response.}
    \label{fig:interferometer}
\end{figure}

The phase response also admits an interferometric interpretation in terms of the separation of the two conditional coherent states. We first consider the undriven reference trajectories, denoted by $\alpha_s^{(0)}(t)$, with $s$ labeling the initial qubit branch throughout the echo sequence. For real $\beta$,
\begin{equation}
    \begin{aligned}
    \alpha_s^{(0)}(t)&=\beta e^{-is\chi f(t)/2},\\
    f(t)&=\begin{cases}
        t, & 0\leq t\leq\tau_0,\\
        2\tau_0-t, & \tau_0<t\leq2\tau_0.
    \end{cases}
    \end{aligned}
\end{equation}
The two reference states have equal real parts and separate along the imaginary quadrature. For $\beta,\chi>0$ and $0<\chi\tau_0\leq\pi$, their separation is
\begin{equation}
    d_0(t)\equiv\operatorname{Im}\alpha_-^{(0)}(t)
    -\operatorname{Im}\alpha_+^{(0)}(t)
    =2\beta\sin\!\left[\frac{\chi f(t)}{2}\right].
\end{equation}
The $\pi$ pulse at $t=\tau_0$ reverses the dispersive rotation, so these undriven trajectories return to the common amplitude $\beta$ at $t=2\tau_0$, as illustrated in gray dashed lines in Fig.~\ref{fig:interferometer}.

The DM drive modifies this closed reference evolution in two related ways. 
First, although the drive acts on both branches with the same instantaneous force, 
the two branches undergo opposite dispersive rotations. The same drive is 
accumulated differently along the two trajectories, so that the echo no longer closes 
perfectly at $t=2\tau_0$. Instead, a residual branch-dependent displacement remains, 
producing the finite opening between the two driven trajectories shown as the colord curves in 
Fig.~\ref{fig:interferometer}.
Second, the same drive also generates a relative phase while the two reference 
trajectories are separated. To first order in $A$, the infinitesimal phase response is 
proportional to the instantaneous branch separation multiplied by the relevant 
quadrature of the DM drive. For the resonant case $\Delta=0$ and $\phi_1=\pi/2$ 
shown in Fig.~\ref{fig:interferometer}, this drive quadrature is constant, giving
\begin{equation}
    |\delta\Phi^{(1)}|
    \propto
    A\int_0^{2\tau_0} d_0(t)\,dt
    \sim
    A\beta\chi\tau_0^2,
    \qquad |\chi\tau_0|\ll1,
\end{equation}
which reproduces the characteristic scaling of Eq.~\ref{dPhi}. 
Geometrically, the phase is proportional to the shaded area swept out by the two arms in the time--quadrature representation like an interferometer.

\begin{figure}
    \centering
    \includegraphics[width=0.99\linewidth]{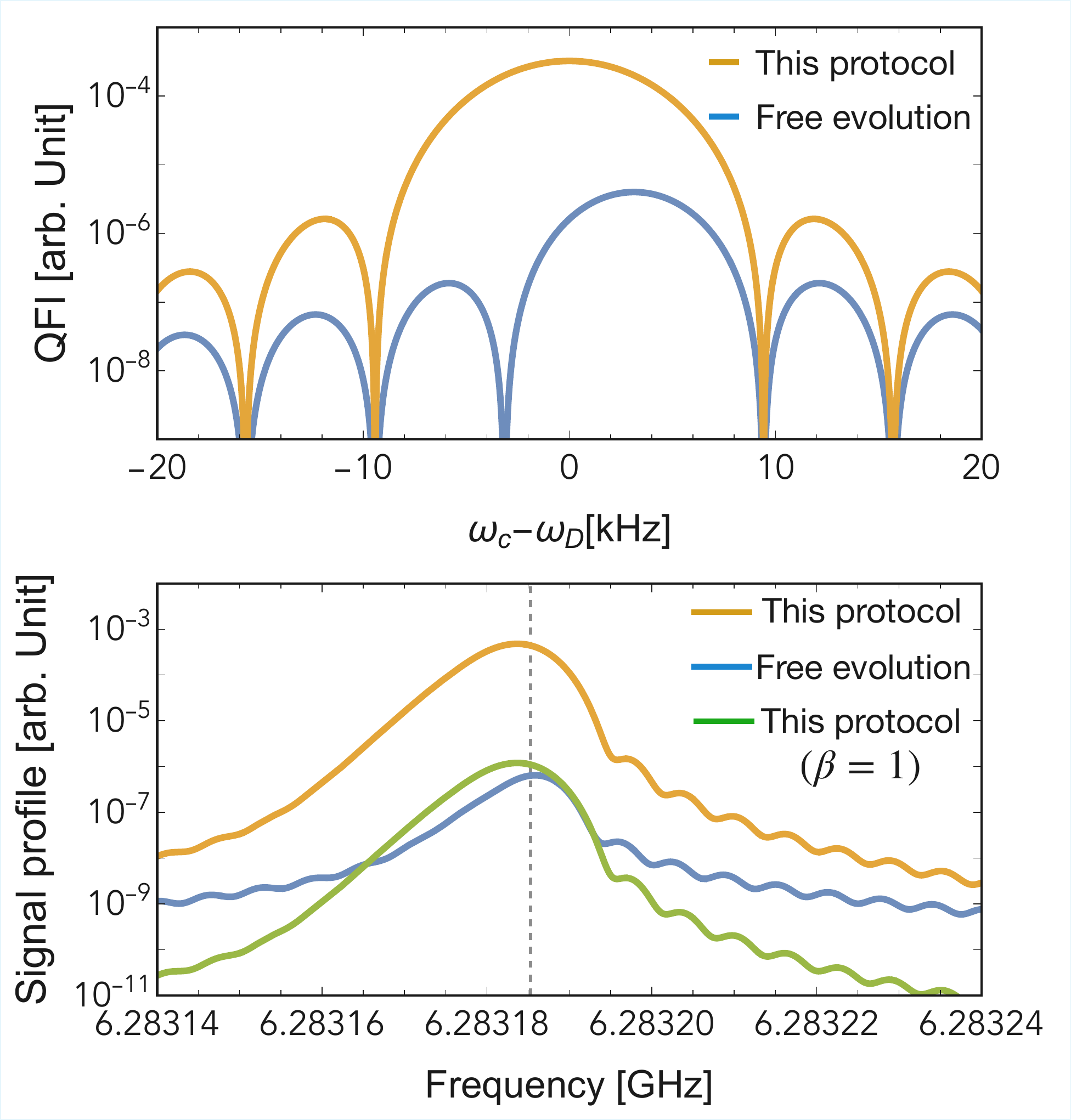}
    \caption{\textit{Upper panel}: Quantum Fisher information versus detuning $\Delta \equiv \omega_c-\omega_D$ of the geometric protocol (yellow) and free evolution (blue) with $\beta = 20$. \textit{Lower panel}: Signal profile for a DM frequency of $1~\mathrm{GHz}$ after convolution with the DM spectrum, shown for $\beta=20$ (yellow), $\beta=1$ (green), and free evolution (blue) with $\chi \tau_0 = \pi$ and cavity frequency $\omega_c=(1+3\times10^{-7})m_{\rm DM}$. 
    }
    \label{fig:QFI}
\end{figure}

Intracavity squeezing up to $14.6\,\mathrm{dB}$, corresponding to $e^r\simeq5.4$, has been demonstrated~\cite{Cai:2025eve}. Combined with pump-induced displacements $\alpha\simeq3$--$5$, this gives $\beta=\alpha e^r\simeq16$--$27$. We therefore retain $\beta=20$ in Fig.~\ref{fig:QFI} as an illustrative benchmark that demonstrates the potential enhancement of the protocol. The projected exclusion curves use the explicitly stated benchmark-dependent values of $\beta$ or the decoherence optimum $\beta_{\rm opt}$.

The spin flip in Block~2 is essential for qubit-cavity DM sensing. The $\hat R$ operator generated by the qubit--cavity interaction in Eq.~\eqref{evo_op} does not commute with the displacement operator $\hat D$. Without the two drift segments, it would effectively alter the squeezing operations in Block $1$ or $3$, preventing the squeezing-induced displacements from canceling. As a result, a residual displacement would remain and overwhelm the DM signal.

Geometric-phase sensing in qubit--oscillator systems, as well as the use of squeezing and displacement followed by an inverse operation to enhance a measurable signal, has been established in Refs.~\cite{Suri:2025hqm,Wang:2026yhs,Cai:2025eve}. This work presents the first practical protocol that incorporates this strategy into DM detection, adapting these operations to a stochastic DM with unknown phase and finite coherence time. We will also quantify the backgrounds and sensitivity in the following sections, 
turning the general geometric-phase sensing concept into a quantitative search strategy tailored to ultralight DM

The measurement of this relative phase can be performed with standard Ramsey interference.
After tracing out the oscillator subspace, a final qubit measurement along the $y$-axis of the Bloch sphere yields a signal
\begin{align}
    S=\frac{\braket{\hat{\sigma}_y}}{2}=- \frac{\delta \Phi}{2} + \mathcal{O}(A^2).
    \label{soperator}
\end{align}
Experimentally, the final qubit pulse $R^{(q)}_x(\pi/2)$ maps $\langle\Op{\sigma}_y\rangle$ onto the excited-state population, which is measured by standard dispersive readout~\cite{PhysRevApplied.17.044016,Kim:2025ywx}. 
The higher-order terms are negligible and discarded in the analysis hereafter.

\prlsection{Enhanced Quantum Fisher Information}{.} 
To quantify the improvement in measurement sensitivity, we evaluate the QFI~\cite{Paris:2008zgg} $\mathcal{F}_Q$ for free evolution and our protocol. The QFI sets the ultimate precision limit for estimating the DM drive amplitude $A$ via the quantum Cram\'er-Rao bound, $(\Delta A)^2 \ge 1/(\nu \mathcal{F}_Q)$ for $\nu$ independent measurements~\cite{Paris:2008zgg}. 
For a pure initial state and an evolution operator of the form $\hat{U}=e^{-i A \hat{H}}$, the QFI is given by the variance of the generator $\hat{H}$ with factor $4$ as a convention:
\begin{equation}
\hspace*{-2mm}
    \mathcal{F}_Q(A) \equiv 4 \left[ \langle \partial_A \Psi | \partial_A \Psi \rangle - |\langle \Psi | \partial_A \Psi \rangle|^2 \right]= 
    4 \text{Var}(\hat H).
\end{equation}
For a standard free evolution, the cavity is initialized in a vacuum state and then displaced by the DM interaction to a coherent state with displacement of Eq.~\eqref{delta_s}. 
To compare with our protocol in an equal-footing way, we consider an evolution of the same duration, $2 \tau_0$. The corresponding QFI up to $\mathcal{O}(A)$ is evaluated to be
\begin{equation}
    \mathcal{F}_{Q,\text{free}} \simeq 16 \tau_0^2 \mathrm{sinc}^2\left(\Omega_{\text{eff},s}\tau_0\right).
    \label{eq:qfi_direct}
\end{equation}
This baseline sensitivity scales purely with the squared evolution time, inherently bounded by standard vacuum fluctuations.

In our protocol, the generators in the cavity and qubit Hilbert spaces are $\hat H_{\rm c}=i(\Sigma\hat a^\dagger-\Sigma^*\hat a)/A$ and $\hat H_{\rm qubit}=\delta\Phi\hat\sigma_z/(2A)$, respectively. For an initial product state of the cavity and qubit, the QFI of the geometric protocol is
\begin{align}
   \mathcal{F}_{Q,{\rm geo}} = 4 [\text{Var}(\hat H_c) + \text{Var}(\hat H_{\text{qubit}})],
\end{align}
it is maximized by the initial state $\frac{1}{\sqrt{2}}(\ket{g}+\ket{e})\otimes\ket{0}$, which gives
\begin{align}
   & \mathcal{F}_{Q,{\rm geo}}^{\rm optimal}\approx
  \dfrac{1}{2} (\alpha e^r\chi \tau_0^2)^2\nonumber\\ &\times\sinc^2\left(\frac{(\Delta+\chi/2)\tau_0}{2}\right) \sinc^2\left(\frac{(\Delta-\chi/2)\tau_0}{2}\right).
  \label{FQI_geo}
\end{align}
after averaging the DM random phase $\phi_1$, $\langle\sin^2\phi_1\rangle=1/2$. 
We here implicitly omit the QFI from cavity space since it does not gain from the geometric phases. The QFI bound can be saturated via Ramsey readout. Background information on the QFI and its detailed derivation can be found in Appendix~\ref{app:QFI}.
The QFI is an ideal precision bound. The projected limits below assume Ramsey readout with spin-projection noise and the decoherence factor $\eta$, and should be interpreted within that noise model.

During the interrogation interval, the cavity state on each qubit branch is a coherent state with mean occupation $\bar n\simeq|\beta|^2$. Therefore, Eq.~\eqref{FQI_geo} scales as $F_{Q,\mathrm{geo}}\propto|\beta|^2\propto\bar n$. Under the conventional photon-number resource counting, this corresponds to SQL scaling. The enhancement discussed here should therefore be understood relative to the conventional vacuum/free-evolution baseline of Eq.~\eqref{eq:qfi_direct}, arising from the larger controlled coherent displacement.

As shown in the upper panel of Fig.~\ref{fig:QFI}, comparison of Eqs.~\eqref{eq:qfi_direct} and \eqref{FQI_geo} gives a QFI enhancement controlled by the single scale $\beta^2=\alpha^2e^{2r}$.
The peak positions of the two differ because Eq.~\eqref{FQI_geo} reaches its maximum at $\Delta = 0$, whereas Eq.~\eqref{eq:qfi_direct} is maximized at $\Delta = \chi/2$. Because of the DM spectral distribution, the signal for a fixed DM mass is spread over frequencies according to $f(\omega_{\rm DM})$. The lower panel of Fig.~\ref{fig:QFI} shows the resulting signal profile after convolution with the DM spectrum. Details are given in Appendix~\ref{singal profile}. Equation~\eqref{FQI_geo} predicts a characteristic double-sinc structure with separation $\chi$. To capture the signal efficiently, the DM linewidth $\delta \omega_{\rm DM} \sim 10^{-6} m_{\rm DM} \sim 2\pi/\tau_{\rm DM}$ should be comparable to or larger than $\chi$, where $\tau_{\rm DM}$ is the DM coherence time. Since the ideal protocol time is around $\tau_0 \sim \tau_{\rm DM}/2$, we choose $\chi\tau_0=\pi$ to maximize the signal. With this choice, the signal of this geometric protocol scales parametrically as $\sim \beta^2$, while the double-$\sinc$ response yields a narrower bandwidth than in the free-evolution protocol.

\begin{figure*}[t!]
    \centering
    \includegraphics[width=0.49\linewidth]{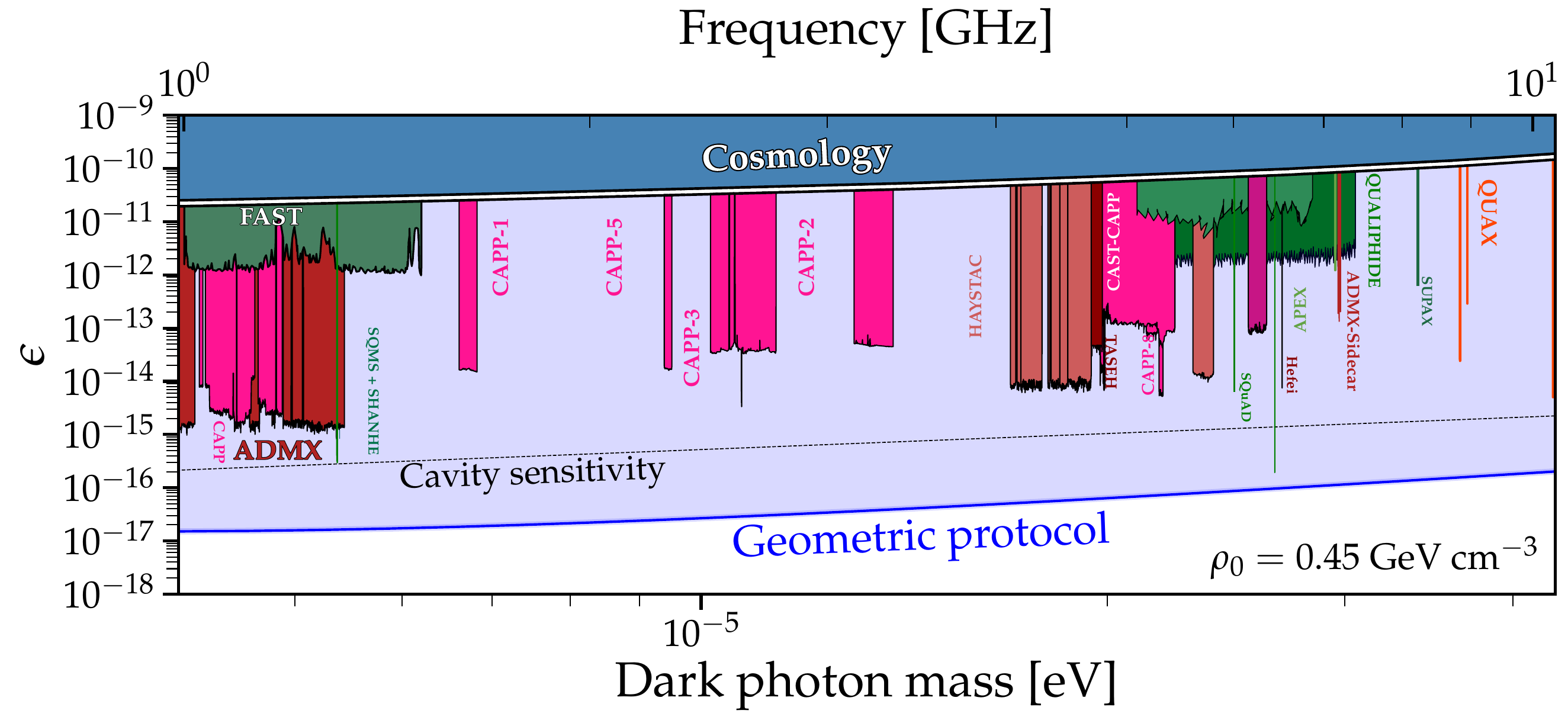}
\includegraphics[width=0.49\linewidth]{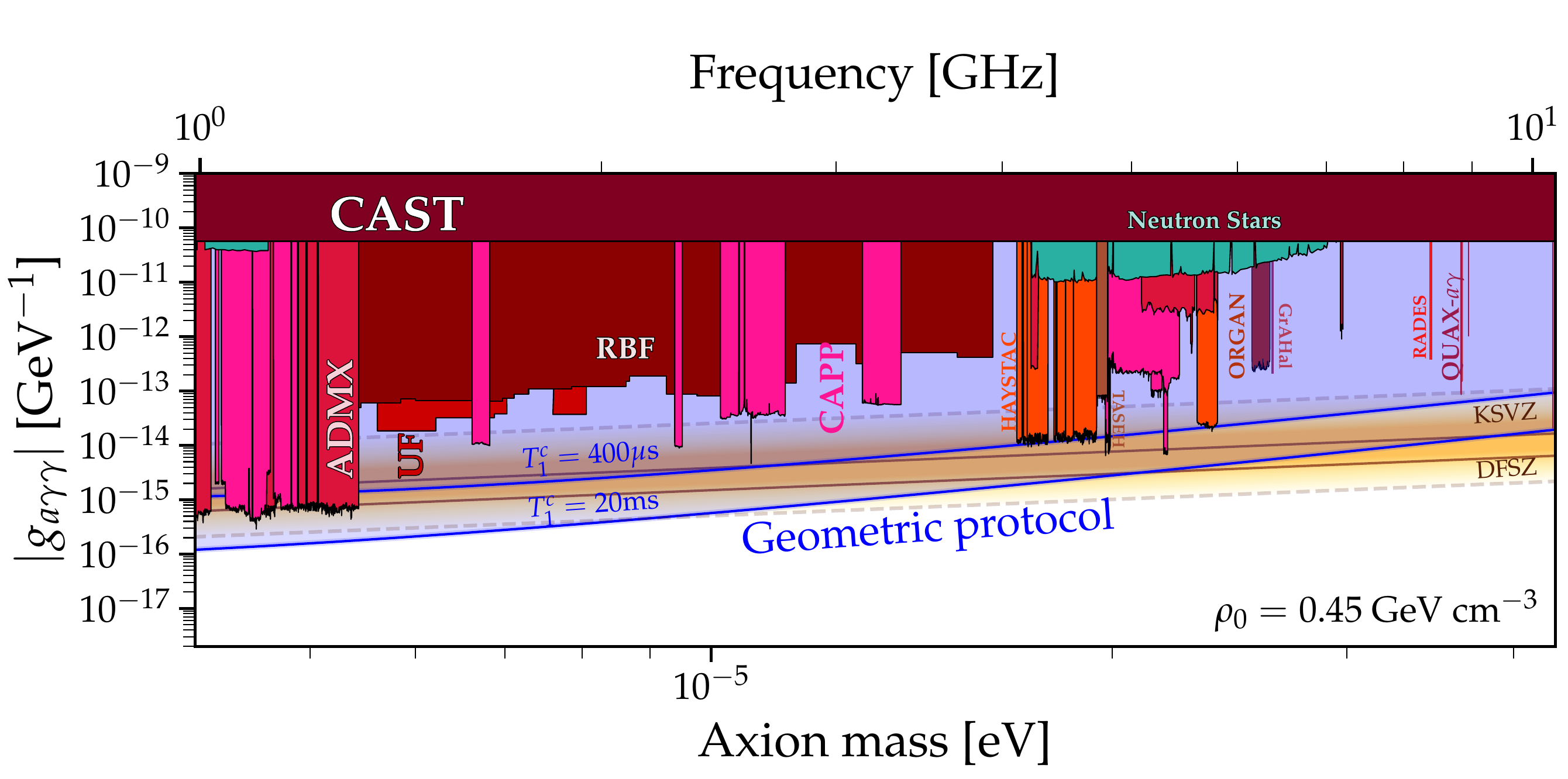}
\caption{\textit{Left}: Projected 95\% C.L. sensitivity to the dark-photon kinetic mixing $\epsilon$ for the geometric protocol (blue solid), compared with existing bounds~\cite{AxionLimits} (colored regions) and the benchmark cavity sensitivity~\cite{SHANHE:2023kxz} (black dotted). 
We take the controlled coherent displacement $\beta$.
The dark-photon displacement is limited by the decoherence optimum $\beta_{\rm opt}$ in Eq.~\eqref{eta}. \textit{Right}: Projected 95\% C.L. sensitivity to the axion-photon coupling $g_{a\gamma\gamma}$ (blue solid), compared with current limits~\cite{AxionLimits} (colored regions). The baseline illustrative and prospective illustrative axion benchmarks use $\beta=1$ and $4.5$, both of which are conditional on a $10\,$T conversion field and an as-yet undemonstrated magnetic-field-compatible cavity--qubit/control architecture. Both panels impose $\chi\tau_0=\pi$ and assume $60\,$s of live integration at each cavity-frequency setting with negligible dead time. The projection lines implicitly assume multiple cavity configurations with corresponding total scan time of 4.4 years.}
    \label{fig:projection}
\end{figure*}

\prlsection{The Impact on Qubit Decoherence}{.}
The geometric protocol amplifies the signal by increasing the qubit-state-dependent separation in cavity phase space, but this also enhances sensitivity to cavity loss, as photon leakage reveals which-path information and reduces the Ramsey contrast.
 This cavity-induced dephasing is an intrinsic measurement back action of the protocol and scales with the phase-space area, $\propto |\beta|^2$.

Considering both the which-path dephasing and intrinsic qubit decoherence, the signal amplitude is suppressed by a factor of
\begin{align}
\eta& =\exp\!\left[-2\tau_0/T_{2,\mathrm{echo}}^{(0)}-\Lambda_\kappa\right],
\label{eta}\nonumber\\
\Lambda_\kappa&\equiv \kappa(1+2\bar n_{\rm th})|\beta|^2\tau_0
\left(1-\sin(\chi\tau_0)/(\chi\tau_0)\right),
\end{align}
as derived in Appendix~\ref{app:background}.
Here, $\kappa\equiv1/T_1^c$ is the inverse cavity lifetime, and $\bar n_{\rm th}$ is the mean thermal photon number in the cavity. The first term describes intrinsic qubit decoherence accumulated during the spin-echo sequence, whereas $\Lambda_\kappa$ captures cavity-induced dephasing from vacuum fluctuations and thermal photons. Since the echo cannot recover information already leaked into the environment, increasing $\beta$ enhances both the signal and the associated dephasing. The optimal operating point $\beta_{\rm opt}=\sqrt{T_1^c/(2\tau_0)}$ for fixed $(\tau_0,T_1^c)$ is therefore determined by this trade-off.

\prlsection{Projected Sensitivity}{.} 
Instead of counting DM-induced cavity excitations, our protocol reads out the signal through qubit Ramsey interferometry. We estimate the reach by repeating the protocol and analyzing the resulting time series in the frequency domain. Each individual measurement yields a binary outcome, $\pm 1$, while the DM information is encoded in the temporal correlations of the data. To account for the finite DM coherence time and the associated stochasticity, we impose a cutoff on signal correlations beyond $\tau_{\rm DM}$, as is typically done in Refs.~\cite{Dror:2022xpi,Chigusa:2024psk}. The resulting power spectral density (PSD) is then obtained from the Wiener--Khinchin theorem.

As shown in Appendix~\ref{app:PSD}, assuming a spin-projection-noise-dominated scenario, the single-qubit projection noise gives a flat background $B_k=\tau_0/2$ for our choice of $\tau=2\tau_0$, whereas the finite DM coherence time generates an excess narrow-band signal PSD $S_k$ centered at the detuning $\Delta$. The width of this spectral feature is set by $\tau_{\rm DM}$, while its amplitude
$\mathcal{S}_k \propto \mathcal{A}
=
8 \eta^2
(\beta \tau_0)^{2}
|A|^{2} / \pi^2$
is proportional to the squared DM-induced drive amplitude, and hence to the squared axion or dark-photon coupling, together with the geometric enhancement and the decoherence factor in Eq.~\eqref{eta}.

We then construct the profile-likelihood test statistic with Asimov data~\cite{Cowan:2010js} for exponentially distributed signal and background, 
\begin{equation}
q=2\sum_k
\left[
\left(1-\frac{B_k}{S_k+B_k}\right)
-
\ln\left(1+\frac{S_k}{B_k}\right)
\right].
\end{equation}
Applying the geometric protocol to dark photon and axion detection and adopting $q=-2.71$ as the $95\%$ exclusion criterion, we obtain the projected sensitivity shown in Fig.~\ref{fig:projection}. In the dark photon case, we choose the cavity coherence time as $T_1^c\sim 300\,$ms and qubit coherence time as $T^{(0)}_{\rm 2,echo}\sim 1\,$ms, which have already been demonstrated in state-of-the-art fluxonium and transmon devices~\cite{Bland:2025vvi,Somoroff:2021elj}. For axion DM searches, we assume a $10\,$T conversion field for illustrative purposes~\cite{Ahn:2021fgb}. Coherent operation of the qubit and Josephson control hardware in an integrated high-field architecture has not been established. HAYSTAC has demonstrated a high-field axion cavity coupled by a microwave chain to spatially separated and magnetically shielded Josephson parametric amplifiers~\cite{HAYSTAC:2020kwv}. This provides a precedent for separated Josephson readout, but not for the coherent dispersive qubit--cavity control required by our protocol. One possible route is therefore to place the qubit and Josephson control hardware in a shielded or spatially separated region and couple it to the conversion cavity through a low-loss microwave link. This architecture and its associated transmission loss, mode matching, and control fidelity remain unvalidated and are not included in our sensitivity calculation.
We present both axion curves as conditional illustrative benchmarks and take $T_1^c=400\,\mu{\rm s}\,(20\,{\rm ms})$ and $\beta=1\,(4.5)$ for the baseline illustrative (prospective illustrative) cases.
In both panels, we set $\chi\tau_0=\pi$ with $\tau_0=\tau_{\rm DM}/2$ and use the effective cavity volume benchmark scaling $V_{\rm eff}=5\,{\rm cm}^3\times(6.44\,{\rm GHz}/(\omega_c/2\pi))^3$. We conservatively take $t_{\rm obs}=60\,$s of integration at each frequency setting.
For the scan-time estimate, we adopt a logarithmic frequency spacing $\delta\nu/\nu=10^{-6}$, comparable to one DM linewidth. At $60\,$s per setting, the accumulated live integration time is approximately $4.4$~yr.
Increasing
$t_{\rm obs}$ increases the number of repeated Ramsey measurements, and for negligible dead time,
$N_{\rm obs}\simeq t_{\rm obs}/(2\tau_0)$, hence improves the
statistical sensitivity. In the experimentally relevant regime
$t_{\rm obs}\gg\tau_{\rm DM}$, the DM field is stochastic across
different coherence intervals, so that the sensitivity to the
dark photon or axion coupling improves approximately as
$t_{\rm obs}^{-1/4}$.
For these benchmark parameters, the geometric protocol improves the sensitivity to dark-photon and axion DM by one to two orders of magnitude. This improvement reflects the enhancement scale $\beta$ produced by the displacement and squeezing operations.

We emphasize that the quoted $60\,\mathrm{s}$ corresponds to a
narrow-band measurement at a single cavity-frequency setting. Covering
the broad DM mass range in Fig.~\ref{fig:projection} requires a frequency scan with multiple distinct cavity configurations, each with a limited tuning range and sufficient overlap between adjacent configurations.
Consequently, the accumulated live integration time is the integration time per point
multiplied by the number of frequency settings.

\prlsection{Conclusion}{.}
We have shown that a dispersively coupled qubit-cavity system can map a weak DM-induced cavity displacement onto a qubit-readable geometric phase. The coherent pump and squeezing set the enhancement through $\beta=\alpha e^r$, while the Ramsey-echo loop supplies the phase-space transduction and rejects the branch-common and dispersive dynamical phases. After including cavity-loss-induced dephasing and spin projection noise, the benchmark model yields projected coupling improvements of up to one to two orders of magnitude. These projections remain conditional on experimental details and, for axions, on magnetic-field-compatible hardware. The protocol provides a concrete route for applying active cQED control to ultralight-DM searches and can be extended to entangled cavities~\cite{Freiman:2025tse}, more general control sequences~\cite{Suri:2025hqm,Wang:2026yhs,Ma:2025orp}, and other quantum-sensing platforms~\cite{Chigusa:2024psk,Aybas:2026rwu,Kennedy:2020bac,Engelhardt:2023qjf}.

\section*{Acknowledgements}

X. M. is supported by the World Premier International Research Center Initiative (WPI), MEXT, Japan (QUP). J. S. is supported by the Japan Society for the Promotion of Science (JSPS) as a
part of the JSPS Postdoctoral Program (Standard) with grant number: P25018, and by the World
Premier International Research Center Initiative (WPI), MEXT, Japan (Kavli IPMU).

\bibliographystyle{utphys}
\bibliography{ref}

\clearpage
\newpage  
\appendix
\onecolumngrid
\renewcommand{\theequation}{S.\arabic{equation}}
\setcounter{equation}{0}

\centerline{\large {Supplemental Material for}}
\medskip

{\centerline{\large \bf{{Enhanced Dark Matter Quantum Sensing via Squeezing-amplified Geometric Phase}}}}
\medskip
{\centerline{Xiaolin Ma, Jie Sheng}}
\bigskip
\bigskip

\appsection{app:DM_coupling}{DM coupling with Cavity}

In this appendix, we derive the coupling of dark-photon and axion DM to the cavity through an effective current.

\subsection*{Dark Photon}

The kinetic mixing between the photon $A$ and dark photon $A'$ can be written as the mass-mixing term
\begin{equation}
\mathcal{L} \supset \epsilon\, m_{A'}^{2}\, A'_{\mu} A^{\mu}.
\end{equation}
This is equivalent to coupling the photon field $A_\mu$ to an effective four-vector current generated by the dark photon background,
\begin{equation}
J_{\mathrm{eff}}^{\mu} = \epsilon\, m_{A'}^{2}\, A'^{\mu}.
\end{equation}
In the Coulomb gauge, the spatial components dominate, and the interaction Hamiltonian is therefore
\begin{equation}
H_{\mathrm{int}}(t)
 = \int_V d^3x\, \bm{A}(\bm{x},t)\cdot \bm{J}_{\mathrm{eff}}(\bm{x},t) 
 = \epsilon\, m_{A'}^{2} \int_V d^3x\, \bm{A}(\bm{x},t)\cdot \bm{A}'(\bm{x},t).
\label{H_int1}
\end{equation}

Retaining a single cavity mode with frequency $\omega_c$, the photon vector potential in the cavity is
\begin{equation}
\bm{A}(\bm{x},t)
= \frac{1}{\sqrt{2\omega_c}}
\left( \hat a^\dagger\, \bm{\epsilon}(\bm{x})\, e^{+i\omega_c t} + \mathrm{h.c.} \right),
\label{SMA}
\end{equation}
with normalization $\int_V d^3x\, \bm{\epsilon}_m^{*}(\bm{x})\cdot \bm{\epsilon}_n(\bm{x}) = \delta_{mn}$.
The dark-photon background can be modeled as a superposition of classical plane waves,
\begin{equation}
\bm{A}'(\bm{x},t) = \sum_{i=1}^{N} \bm{A}'_{i}(\bm{x},t),
\end{equation}
with each component
\begin{equation}
\bm{A}'_{i}(\bm{x},t)
= A'_{i,0}\, \bm{\epsilon}_i \cos\!\left(\omega_i t - \bm{k}_i\cdot\bm{x} + \phi_i\right).
\label{darkA}
\end{equation}
Assuming the dark photon background forms a single coherent component ($N=1$), the amplitude $A'_0$ with no index $i$ is fixed by the local DM energy density,
\begin{equation}
\rho_{\mathrm{DM}} \simeq \frac{m_{A'}^{2}(A'_{0})^{2}}{2}\,, \quad A'_{0} \simeq \sqrt{\frac{2\rho_{\mathrm{DM}}}{m_{A'}^{2}}}
= \frac{\sqrt{2\rho_{\mathrm{DM}}}}{m_{A'}}.
\end{equation}
In such a case, the dark photon field frequency is $\omega_D$ and ${\bm k}$ is the corresponding dark photon momentum. It also has a random phase $\phi$.

Substituting $\bm{A}$ in Eq.~\eqref{SMA} and $\bm{A}'$ in Eq.~\eqref{darkA} into $H_{\mathrm{int}}$ Eq.~\eqref{H_int1} and
Expanding the cosine into exponentials yields both slowly-rotating terms $\propto e^{\pm i(\omega_c-\omega_D)t}$ and fast-rotating terms $\propto e^{\pm i(\omega_c+\omega_D)t}$ where $\omega_D$ is the dark photon frequency of the coherent mode.
Under the rotating-wave approximation (RWA), keeping only the near-resonant $\omega_c - \omega_D$ terms, one obtains
\begin{equation}
H_{\mathrm{int}}(t)
= \sum_{i}
\left[
\epsilon m_{A'}^{2} A'_{0}\sqrt{\frac{V_{\mathrm{eff}}}{2\omega_c}}\,
\hat a^\dagger\, e^{i(\omega_c-\omega_D)t-i\phi_i}
+ \mathrm{h.c.}
\right].
\label{hint_dp_1}
\end{equation}
The effective overlap volume is defined as
\begin{equation}
V_{\mathrm{eff}}
\equiv
\left|
\int_V d^3x\,
\bm{\epsilon}(\bm{x})\cdot \bm{\epsilon}_D\, e^{-i\bm{k} \cdot\bm{x}}
\right|^{2}.
\end{equation}
Here, $\bm{\epsilon}_D$ is the polarization vector of the coherent dark-photon mode.

Defining $\Delta\equiv\omega_c-\omega_D$, the interaction Hamiltonian takes the form used in the main text,
\begin{equation}
H_{\mathrm{int}}(t)=
A\, \hat a^\dagger e^{i(\Delta t-\phi_1)} + \mathrm{h.c.},
\end{equation}
Comparison with Eq.~\eqref{hint_dp_1} gives the drive amplitude
\begin{equation}
A \equiv \epsilon\, m_{A'}\, \sqrt{\rho_{\mathrm{DM}}}\,\sqrt{\frac{V_{\mathrm{eff}}}{\omega_c}}.
\end{equation}

\subsection{Axion}

In the presence of an external magnetic field, the axion can also induce a similar drive on the cavity. The axion--photon interaction is
\begin{equation}
\mathcal{L}_{a\gamma\gamma}
= -\frac{g_{a\gamma\gamma}}{4}a\,F_{\mu\nu}\tilde F^{\mu\nu}
= g_{a\gamma\gamma}\, a\, \bm{E}\cdot \bm{B}.
\end{equation}
We decompose the electromagnetic field into the static applied field $\bm B_0$ and the cavity fields $\bm B_{\rm cav}$ and $\bm E_{\rm cav}$,
\begin{equation}
\bm{B} = \bm{B}_0 + \bm{B}_{\mathrm{cav}},
\qquad
\bm{E}=\bm{E}_{\mathrm{cav}},
\end{equation}
and retain the term linear in the cavity field,
\begin{equation}
\mathcal{L}_{\mathrm{int}} \simeq g_{a\gamma\gamma}\, a\, \bm{E}_{\mathrm{cav}}\cdot \bm{B}_0.
\end{equation}
The corresponding interaction Hamiltonian is
\begin{equation}
H_{\mathrm{int}}(t)
\simeq -g_{a\gamma\gamma}\int_V d^3x\,
a(t,\bm{x})\, \bm{E}_{\mathrm{cav}}(\bm{x},t)\cdot \bm{B}_0(\bm{x}).
\end{equation}

Keeping one cavity mode with frequency $\omega_c$,
\begin{equation}
\bm{E}_{\mathrm{cav}}(\bm{x},t)
= \bm{E}_{\mathrm{zpf}}(\bm{x})
\left(\hat a\,e^{-i\omega_c t} + \hat a^\dagger e^{+i\omega_c t}\right),
\end{equation}
with the zero-point field written as
\begin{equation}
\bm{E}_{\mathrm{zpf}}(\bm{x})=\sqrt{\frac{\omega_c}{2}}\bm{\epsilon}(\bm{x}),
\qquad
\int_V d^3x\,|\bm{\epsilon}(\bm{x})|^2=1.
\end{equation}
The interaction Hamiltonian $H_{\mathrm{int}}$ becomes
\begin{equation}
H_{\mathrm{int}}(t)
 =
-g_{a\gamma\gamma}\sqrt{\frac{\omega_c}{2}}
\left(\hat a\,e^{-i\omega_c t} + \hat a^\dagger e^{+i\omega_c t}\right) 
\int_V d^3x\, a(t,\bm{x})\,\bm{\epsilon}(\bm{x})\cdot \bm{B}_0(\bm{x}).
\end{equation}

Similarly, the axion field can be modeled as a classical oscillating wave,
\begin{equation}
a(t,\bm{x})
= a_0\cos(\omega_a t-\bm{k}\cdot\bm{x}+\phi_a),
\end{equation}
with $\omega_a\simeq m_a$ and $|\bm{k}|\simeq m_a v$. Expanding the cosine, applying the RWA, and retaining the near-resonant terms give the same Hamiltonian form as in the dark-photon case,
\begin{equation}
H_{\mathrm{int}}(t)
\approx
A\,\hat a^\dagger e^{i\left((\omega_c-\omega_a)t-\phi_a\right)} + \mathrm{h.c.},
\end{equation}
where the detuning is $\Delta\equiv\omega_c-\omega_a$.
The effective axion drive amplitude is
\begin{equation}
A
=
\frac{g_{a\gamma\gamma}}{2}\,a_0\,\sqrt{\frac{\omega_c}{2}}\,
\left|\int_V d^3x\, \bm{\epsilon}(\bm{x})\cdot \bm{B}_0(\bm{x})\,e^{-i\bm{k}\cdot\bm{x}}\right|.
\end{equation}

It is common to rewrite the overlap in terms of a magnetic-field strength $B_0$, a mode form factor $C$, and an effective volume $V$. One standard definition of the form factor is
\begin{equation}
C
\equiv
\frac{\left|\int_V d^3x\, \bm{E}_{\mathrm{mode}}(\bm{x})\cdot \bm{B}_0(\bm{x})\right|^2}
{\left(\int_V d^3x\,|\bm{E}_{\mathrm{mode}}(\bm{x})|^2\right)\left(\int_V d^3x\,|\bm{B}_0(\bm{x})|^2\right)}.
\end{equation}
Under the usual assumptions of approximately uniform $\bm{B}_0$ and $kL\ll 1$, one may estimate
\begin{equation}
A \sim g_{a\gamma\gamma}\, a_0\, B_0\, \sqrt{\frac{\omega_c C V}{2}},
\end{equation}
up to an $\mathcal{O}(1)$ factor depending on field normalization conventions and $C \simeq 1$. Finally, the axion field amplitude is related to the local axion energy density by
\begin{equation}
\rho_a \simeq \frac{1}{2} m_a^2 a_0^2, \quad 
a_0=\frac{\sqrt{2\rho_a}}{m_a}.
\end{equation}
Substituting into the estimate for $A$ gives
\begin{equation}
A 
= g_{a\gamma\gamma}\,B_0\,\sqrt{\rho_a}\,\sqrt{\frac{\omega_c C V}{m_a^2}}.
\end{equation}

\appsectionnolabel{\texorpdfstring{Derivation of the Evolution under the Dark Matter Drive}{Derivation of the Evolution under the Dark Matter Drive}}
\label{app:EvolutionOp}

In this appendix, we present the Magnus expansion in the interaction picture corresponding to the Hamiltonian in Eq.~\eqref{hamiltonian}, and derive the full time-evolution operator of our system.
We first split the system Hamiltonian $H_I(t) \equiv H_0 + V(t)$ into a time-independent free evolution part $H_0$ and a time-dependent DM driving part $V(t)$:
\begin{equation}
    H_0 = \frac{s\chi}{2} \adag \aop = \Omega_s \adag \aop\,, \quad V(t) = A \left( \adag e^{i(\Delta t - \phi_1)} + \aop e^{-i(\Delta t - \phi_1)} \right)
\end{equation}
Using the evolution operator 
$U_0(t) = e^{-i H_0 t}$ generated by $H_0$, we can transform the DM interaction term $V(t)$ into the interaction picture as
\begin{equation}
    \tilde{H}_I(t) = U_0^\dagger(t) V(t) U_0(t) \nonumber 
    = A \left( \adag e^{i(\Omega_{\text{eff}, s} t - \phi_1)} + \aop e^{-i(\Omega_{\text{eff}, s} t - \phi_1)} \right),
\end{equation}
where $\Omega_{\text{eff}, s} \equiv \Omega_s + \Delta$ is the effective detuning.

Using the Magnus expansion, the time evolution operator in the interaction picture can be written as $U_I(t_f, t_i) = \exp \left( \Omega_1 + \Omega_2 + \dots \right)$. The first-order Magnus term is
\begin{equation}
    \Omega_1(t_i,t_f) = -i \int_{t_i}^{t_f} \tilde{H}_I(t') dt' = \delta_s \adag - \delta_s^* \aop.
\end{equation}
Defining $\tau\equiv t_f-t_i$, this term yields the displacement operator $\hat D(\delta_s)\equiv\exp\Omega_1$ with amplitude $\delta_s=-\frac{A e^{-i\phi_1}}{\Omega_{\text{eff},s}}e^{i\Omega_{\text{eff},s}t_f}\left[1-e^{-i\Omega_{\text{eff},s}\tau}\right]$, in agreement with the main text.
Using the quadrature operators $\hat x$ and $\hat p$,
$
\hat x=\frac{\hat a+\hat a^\dagger}{\sqrt{2}},
\hat p=\frac{\hat a-\hat a^\dagger}{i\sqrt{2}}$,
the displacement operator can be written as
\begin{equation}
\hat D(\delta)=\exp\!\left(-i\sqrt{2}\,(\mathrm{Re}\,\delta)\,\hat p
+i\sqrt{2}\,(\mathrm{Im}\,\delta)\,\hat x\right).
\end{equation}
The real and imaginary parts of $\delta$ therefore generate shifts in position and momentum, respectively,
\begin{equation}
\hat D^\dagger(\delta)\,\hat x\,\hat D(\delta)=\hat x+\sqrt{2}\,\mathrm{Re}\,\delta,\qquad
\hat D^\dagger(\delta)\,\hat p\,\hat D(\delta)=\hat p+\sqrt{2}\,\mathrm{Im}\,\delta.
\end{equation}
Thus, $\delta$ is the displacement vector in phase space,
\begin{equation}
\delta=\frac{\Delta x+i\Delta p}{\sqrt{2}}.
\end{equation}

The second-order term arises from the non-commutativity of the interaction Hamiltonian at different times, yielding a strictly $c$-number phase:
\begin{equation}
    \Omega_2 = \frac{1}{2} \int_{t_i}^{t_f} dt_1 \int_{t_i}^{t_1} dt_2 [\tilde{H}_I(t_1), \tilde{H}_I(t_2)] \nonumber 
    = -i \frac{A^2}{\Omega_{\text{eff}, s}} \left[ \tau - \frac{\sin(\Omega_{\text{eff}, s} \tau)}{\Omega_{\text{eff}, s}} \right] \equiv i \Phi_G(s).
\end{equation}
This term $\Phi_G(s)$ represents a spin-dependent geometric phase. Since it is a pure number without operators, it can commute with other operators thus the Magnus expansion stops at this second order. The evolution operator in interaction picture is therefore $U_I(t_i,t_f) = e^{i \Phi_G(s)} \hat D(\delta_s)$.

\appsection{app:protocol}{\texorpdfstring{Details of the Evolution under the Geometric Protocol}{Details of the Evolution under the Geometric Protocol}}
In this Appendix, we provide a detailed derivation of the evolution operator for each step of our geometric protocol, the total evolution operator, and the resulting overall geometric phase.

\subsection{\texorpdfstring{Commutation Relations}{Commutation Relations}}
To begin with, we first introduce several general commutation relations that will be used throughout the derivation.

\noindent \textbf{1. Rotation and Displacement Operators}:
The rotation operator is generated by the number operator $\hat N=\hat a^\dagger\hat a$,
\begin{equation}
\hat{R} \equiv e^{-i s \chi t\, \hat{a}^\dagger \hat{a}/2}.
\end{equation}
With the standard identities,
\begin{equation}
e^{\lambda \hat{N}} \hat{a}\, e^{-\lambda \hat{N}} = e^{-\lambda}\hat{a}\,, \quad 
e^{\lambda \hat{N}} \hat{a}^\dagger e^{-\lambda \hat{N}} = e^{+\lambda}\hat{a}^\dagger
\end{equation}
Conjugating the exponent of the displacement operator by $\hat R$ gives
\begin{equation}
\hat{R}^\dagger \left(\alpha \hat{a}^\dagger - \alpha^* \hat{a}\right)\hat{R}
=
\alpha e^{i s \chi t/2}\hat{a}^\dagger
-
\alpha^* e^{-i s \chi t/2}\hat{a}.
\end{equation}
Therefore, conjugation by $\hat R$ rotates the displacement amplitude,
\begin{equation}
\hat{R}^\dagger \hat{D}(\alpha)\hat{R}
=
\exp\left(
\alpha e^{i s \chi t/2}\hat{a}^\dagger
-
\alpha^* e^{-i s \chi t/2}\hat{a}
\right)
=
\hat{D}\!\left(\alpha e^{i s \chi t/2}\right).
\label{RD}
\end{equation}

\noindent \textbf{2. Squeezing and Displacement Operators}:
The single-mode squeezing operator is defined as
\begin{equation}
\hat{S}(r)
\equiv 
\exp\left[
\frac{r}{2}\left(\hat{a}^2-\hat{a}^{\dagger 2}\right)
\right],
\end{equation}
Its action on the annihilation and creation operators is
\begin{equation}
\hat{S}^\dagger(r)\hat{a}\hat{S}(r)
=
\hat{a}\cosh r - \hat{a}^\dagger \sinh r\,, \quad
\hat{S}^\dagger(r)\hat{a}^\dagger \hat{S}(r)
=
\hat{a}^\dagger \cosh r - \hat{a}\sinh r.
\end{equation}
Similarly, its action on the exponent of the displacement operator gives
\begin{equation}
\begin{split}
    \hat{S}^\dagger(r)
\left(\alpha \hat{a}^\dagger-\alpha^*\hat{a}\right)
\hat{S}(r)
& =
\alpha\left(\hat{a}^\dagger\cosh r-\hat{a}\sinh r\right)
-
\alpha^*\left(\hat{a}\cosh r-\hat{a}^\dagger\sinh r\right)\\
& =
\left(\alpha\cosh r+\alpha^*\sinh r\right)\hat{a}^\dagger
-
\left(\alpha^*\cosh r+\alpha\sinh r\right)\hat{a}.
\end{split}
\end{equation}
This shows that, in general,
\begin{equation}
\hat{S}^\dagger(r)\hat{D}(\alpha)\hat{S}(r)
=
\hat{D}(\mu\alpha+\nu\alpha^*),
\end{equation}
with $\mu=\cosh r,\qquad \nu=\sinh r$.
If a squeezing phase is included, the transformation is more general. In the present setup, we take a real displacement along the $x$ direction, namely
$\alpha \in \mathbb{R}$.
Therefore, 
\begin{equation}
\mu\alpha+\nu\alpha^*
=
(\cosh r+\sinh r)\alpha
=
e^r \alpha.
\end{equation}
The squeezing operation increases the displacement magnitude according to
\begin{equation}
\hat{S}^\dagger(r)\hat{D}(\alpha)\hat{S}(r)
=
\hat{D}(\alpha e^r),
\label{SD}
\end{equation}

\noindent \textbf{3. Products of Displacement Operators}:
We define the displacement operator and its exponent by
\begin{equation}
\hat{D}(\alpha)=\exp\left(\alpha \hat{a}^\dagger-\alpha^*\hat{a}\right)
\end{equation}
as,
\begin{equation}
\hat{X}(\alpha)=\alpha \hat{a}^\dagger-\alpha^*\hat{a}.
\end{equation}
Using the commutation relation between the annihilation and creation operators gives
\begin{equation}
[\hat{X}(\alpha),\hat{X}(\beta)]
=
\alpha\beta^*[\hat{a}^\dagger,-\hat{a}]
-
\alpha^*\beta[\hat{a},\hat{a}^\dagger]
=
\alpha\beta^*-\alpha^*\beta,
\end{equation}
which is a $c$-number, so it commutes with both $\hat{a}$ and $\hat{a}^\dagger$. Therefore, the Baker--Campbell--Hausdorff formula truncates as
\begin{equation}
e^{\hat{X}(\alpha)}e^{\hat{X}(\beta)}
=
e^{\hat{X}(\alpha)+\hat{X}(\beta)+\frac{1}{2}[\hat{X}(\alpha),\hat{X}(\beta)]}.
\end{equation}

Therefore, the standard multiplication rule for displacement operators is
\begin{equation}
\hat{D}(\alpha)\hat{D}(\beta)
=
\exp\left(
\frac{\alpha\beta^*-\alpha^*\beta}{2}
\right)
\hat{D}(\alpha+\beta).
\end{equation}
With 
\begin{equation}
\frac{\alpha\beta^*-\alpha^*\beta}{2}
=
\frac{2i\,\mathrm{Im}(\alpha\beta^*)}{2}
=
i\,\mathrm{Im}(\alpha\beta^*),
\end{equation}
Finally, the product of two displacement operators is
\begin{equation}
\hat D(\alpha)\hat D(\beta)
=
e^{\,i\,\mathrm{Im}(\alpha\beta^*)}\,
\hat D(\alpha+\beta).
\label{DD}
\end{equation}

\subsection{Geometric Protocol and Total Phase}

\begin{enumerate}

    \item \textbf{Block 1 ($U_1$)}: Application of a strong pump and squeezing operator $U_1 = \hat{S}^\dagger(r)\hat{D}(\alpha)\hat{S}(r) = \D(\alpha e^r)$ at time $t=0$.

    \item \textbf{Block 2 ($U_2$)}: Evolution over $t \in [0, 2\tau_0]$. 
    This block involves a free evolution of the cavity for $\tau_0$ under spin state $s$ (accumulating drift $\delta_{1a}$), followed by a qubit $\pi$-pulse (spin flip), and another evolution for $\tau_0$ under spin state $-s$ (accumulating drift $\delta_{1b}$).
In the full qubit-cavity Hilbert space, this block is $U(\tau_0,2\tau_0)R_x^{(q)}(\pi)U(0,\tau_0)$. The expression below is branch resolved, and the physical qubit pulse is represented by changing the conditional label from $s$ to $-s$.
\begin{align}
    U_2 &= [e^{i \Phi_G(-s)}\R^\dagger(s) \D(\delta_{1b})] [e^{i \Phi_G(s)}\R(s) \D(\delta_{1a})] \nonumber \\
    &= e^{i (\Phi_G(-s) + \Phi_G(s))} \D(\delta_{1b} e^{i\phi}) \D(\delta_{1a}).
\end{align}
At the second step, we have applied the relationship Eq.~\eqref{RD} and thus 
the frame rotation phase is $\phi = s \chi \tau_0 / 2$. Omitting the global geometric phase which is symmetric across spin states $\Phi_G(-s) + \Phi_G(s)$, the Block 2 operator can be further simplified to
\begin{equation}
    U_2 = e^{i \Phi_{\rm disp}} \D(\Sigma).
\end{equation}
by applying Eq.~\eqref{DD}. The effective displacement is $\Sigma = \delta_{1b} e^{i\phi} + \delta_{1a}$ and phase is $\Phi_{G} = \Impart(\delta_{1b}e^{i \phi}\delta_{1a}^*)$.
    
    \item \textbf{Block 3 ($U_3$)}: The final operation $U_3=\D(-\alpha e^r)$ cancels the controlled displacement generated in Block~1.
\end{enumerate}

The measurable signal emerges from the final state of the oscillator after the full sequence $U_{\text{tot}} = U_3 U_2 U_1$. Defining the amplified pump amplitude as $\beta = \alpha e^r$, we have:
\begin{equation}
    U_{\text{tot}} = e^{i \Phi_{\rm disp}}\D(-\beta) \D(\Sigma) \D(\beta).
\end{equation}
Repeated application of Eq.~\eqref{DD} combines the three displacement operators into a single displacement and an overall phase factor.
\begin{align} 
    U_{\text{tot}} &= \D(\Sigma) e^{i (2\beta \Impart(\Sigma)+\mathcal{O}(A^2))},
\end{align}
The simplified effective displacement is

\begin{align}
   \Sigma= \frac{-A e^{i(\Delta\tau_0 - \phi_1)}}{\Delta^2 - \chi^2/4} \left[ 2i\Delta \sin(\Delta\tau_0) + s\chi \cos(\Delta\tau_0) - s\chi e^{is\chi\tau_0/2} \right]
\end{align}

For real $\beta$ and a DM-induced displacement much smaller than the controlled displacement $|\beta|$, the accumulated signal phase is
\begin{equation}
    \Phi_{\text{total}} \approx 2 \beta \Impart(\Sigma).
\end{equation}

Using Euler identities, we define the single-interval drift coefficient
\begin{equation}
    C_0 (s) = -2i \frac{A}{\Omega_{\text{eff},s}} \sin\left(\frac{\Omega_{\text{eff},s}\tau_0}{2}\right) e^{i(\frac{\Omega_{\text{eff},s}\tau_0}{2} - \phi_1)}.
\end{equation}
The total phase on a given spin branch is
\begin{equation}
    \Phi_{\text{total}} = 2\alpha e^r \Impart\left[e^{i2\Delta \tau_0}(C_0(s)+C_0(-s)e^{i \Omega_{\text{eff},s}\tau_0})\right].
\end{equation}
The factor $e^{i2\Delta\tau_0}$ is the DM-drive phase advanced to the final recombination time $t=2\tau_0$. It follows from the factors $e^{i\Delta t_i}$ in the two drift amplitudes in Eq.~\eqref{delta_s} and is not an additional measurable phase.

In a Ramsey measurement, the observable is the relative phase between the two qubit branches. The phase shift $\delta\Phi=\Phi_{+1}-\Phi_{-1}$ is
\begin{align}
    \delta\Phi = A\alpha e^r\chi \tau_0^2\sin(\Delta \tau_0-\phi_1) \sinc\left(\frac{(\Delta+\chi/2)\tau_0}{2}\right) \sinc\left(\frac{(\Delta-\chi/2)\tau_0}{2}\right).
\end{align}
The geometric signal is enhanced by the controlled-displacement scale $\alpha e^r$ relative to the bare DM-induced displacement.

\phantomsection
\label{app:reversal}
\noindent\textbf{Calibration of imperfect displacement reversal.}
Let the implemented final cavity displacement on qubit branch $s=\pm1$ be $\hat D(-\beta+\varepsilon_s)$, where $\beta$ is taken to be real. Applying Eq.~\eqref{DD} gives
\begin{align}
\hat D(-\beta+\varepsilon_s)\hat D(\Sigma_s)\hat D(\beta)
    &=
    e^{i\Phi_s^{(\varepsilon)}}\hat D(\Sigma_s+\varepsilon_s),
    \nonumber\\
 \Phi_s^{(\varepsilon)}
    &=
    2\beta\,\operatorname{Im}\Sigma_s
    +\beta\,\operatorname{Im}\varepsilon_s
    +\operatorname{Im}(\varepsilon_s\Sigma_s^*).
\end{align}
The first term is the intended geometric phase. The additional branch-differential phase is
\begin{align}
    \delta\Phi_\varepsilon
    &=
  \beta\,\operatorname{Im}(\varepsilon_+-\varepsilon_-)
    +\operatorname{Im}(\varepsilon_+\Sigma_+^*
    -\varepsilon_-\Sigma_-^*).
\end{align}
A branch-independent and reproducible residual displacement is monitored with pump-off or far-detuned reference shots and removed as an offset. The qubit-correlated or cycle-to-cycle component should satisfy the condition $|\delta\Phi_\varepsilon|\ll|\delta\Phi_{\rm DM}|$. In the experimentally relevant regime $|\Sigma_s|\ll|\beta|$, the leading requirement is $|\operatorname{Im}(\varepsilon_+-\varepsilon_-)|\ll|\delta\Phi_{\rm DM}|/|\beta|$. This signal-dependent inequality is the appropriate reversal tolerance.

\appsectionnolabel{Details Concerning the Quantum Fisher Information}
\label{app:QFI}

In this appendix, we define the QFI and derive it for free and geometric evolution. For a pure state $|\Psi(A)\rangle$ that depends on the parameter $A$, an infinitesimal shift $A\to A+dA$ changes the state according to
\begin{equation}
|\Psi(A+dA)\rangle
=
|\Psi(A)\rangle
+dA\,\partial_A |\Psi\rangle
+\frac{dA^2}{2}\,\partial_A^2 |\Psi\rangle+\cdots .
\end{equation}
Its inner product with the initial state is
\begin{equation}
\langle \Psi(A)|\Psi(A+dA)\rangle
=
1+dA\,\langle \Psi|\partial_A\Psi\rangle
+\frac{dA^2}{2}\,\langle \Psi|\partial_A^2\Psi\rangle+\cdots .
\label{inner}
\end{equation}
The normalization condition is $\langle\Psi|\Psi\rangle=1$.
Differentiating Eq.~\eqref{inner} with respect to $A$ gives $\langle \partial_A \Psi | \Psi \rangle + \langle \Psi | \partial_A \Psi \rangle = 0$.
Therefore, $\langle \Psi | \partial_A \Psi \rangle$
must be purely imaginary.
Expanding the squared overlap to second order, one obtains
\begin{equation}
|\langle \Psi(A)|\Psi(A+dA)\rangle|^2
=
1-dA^2\left(
\langle \partial_A \Psi | \partial_A \Psi \rangle
-
|\langle \Psi | \partial_A \Psi \rangle|^2
\right)
+\mathcal{O}(dA^3).
\end{equation}
By definition, the QFI is four times the coefficient of this second-order ``distance'' term:
\begin{equation}
\mathcal{F}_Q(A)
\equiv 
4\left(
\langle \partial_A \Psi | \partial_A \Psi \rangle
-
|\langle \Psi | \partial_A \Psi \rangle|^2
\right).
\label{QFI_f}
\end{equation}
Its physical meaning is that when the parameter is changed slightly, the quantum state moves through Hilbert space; the farther it moves, the larger the QFI.

If the parameter enters the state through a unitary evolution,
\begin{equation}
|\psi(A)\rangle = e^{-iA\hat H}|\psi_0\rangle, \quad \partial_A |\psi(A)\rangle = -i\hat H |\psi(A)\rangle,
\label{UiaH}
\end{equation}
one can substitute Eq.~\eqref{UiaH} into the pure-state QFI formula Eq.~\eqref{QFI_f}, 
\begin{equation}
\langle \partial_A \psi | \partial_A \psi \rangle
=
\langle \psi | \hat H^2 | \psi \rangle, \quad \langle \psi | \partial_A \psi \rangle
=
-i\langle \hat H \rangle.
\end{equation}
Therefore, the QFI is four times the variance of the generator $\hat H$,
\begin{equation}
\mathcal{F}_Q
=
4\left(\langle \hat H^2 \rangle - \langle \hat H \rangle^2\right)
=
4\,\mathrm{Var}(\hat H).
\end{equation}

We next consider a coherent state $|\Psi(A)\rangle = |\beta(A)\rangle$ with phase-space displacement $\beta (A)$. Its standard representation is
\begin{equation}
|\beta\rangle
=
e^{-|\beta|^2/2}e^{\beta \hat a^\dagger}|0\rangle.
\end{equation}
Differentiating it with respect to the parameter $A$ and using the chain rule gives
\begin{equation}
\partial_A |\beta\rangle
=
\left[
(\partial_A \beta)\hat a^\dagger
-
\frac{1}{2}
\left(
\beta^* \partial_A \beta + \beta \,\partial_A \beta^*
\right)
\right]|\beta\rangle.
\end{equation}
The inner product becomes
\begin{equation}
\langle \beta | \partial_A \beta \rangle
=
\frac{1}{2}
\left(
\beta^* \partial_A \beta
-
\beta\, \partial_A \beta^*
\right).
\end{equation}
Using the standard coherent-state relations
\begin{equation}
\hat a |\beta\rangle = \beta |\beta\rangle, \quad \langle \beta | \hat a^\dagger = \beta^* \langle \beta |,
\end{equation}
gives
\begin{equation}
\langle \partial_A \beta | \partial_A \beta \rangle
=
|\partial_A \beta|^2
+
\frac{1}{4}
\left(
\beta^* \partial_A \beta
-
\beta\, \partial_A \beta^*
\right)^2.
\end{equation}

According to the QFI,
\begin{equation}
\mathcal{F}_Q
=
4\left(
\langle \partial_A \beta | \partial_A \beta \rangle
-
|\langle \beta | \partial_A \beta \rangle|^2
\right),
\end{equation}
the two terms involving $\beta$ itself cancel exactly, leaving
\begin{equation}
\mathcal{F}_Q = 4 |\partial_A \beta|^2.
\end{equation}
This result has a direct phase-space interpretation. The parameter $A$ moves the state through phase space with speed $|\partial_A\beta|$, and the QFI is proportional to the square of this speed.

\textbf{QFI for free evolution.}
As shown in the main text, free evolution under the DM field generates the displacement
\begin{equation}
\delta_s(t_i,t_f)
=
-\frac{A e^{-i\phi_1}}{\Omega_{\mathrm{eff},s}}
e^{i\Omega_{\mathrm{eff},s}\tau}
\left(1-e^{-i\Omega_{\mathrm{eff},s}\tau}\right)
e^{i\Delta t_i}.
\end{equation}
For an evolution time $2\tau_0$, the displacement is
\begin{equation}
\beta(A)=\delta_s(0,2\tau_0) =-\frac{A e^{-i\phi_1}}{\Omega_{\mathrm{eff},s}}
e^{i2\Omega_{\mathrm{eff},s}\tau_0}
\left(1-e^{-i2\Omega_{\mathrm{eff},s}\tau_0}\right).
\label{betaA_f}
\end{equation}
Using the identity
\begin{equation}
1-e^{-i2x}=2i\,e^{-ix}\sin x,
\end{equation}
Eq.~\eqref{betaA_f} can be rewritten as
\begin{equation}
\beta(A)
=
-2iA\,e^{-i\phi_1}
e^{i\Omega_{\mathrm{eff},s}\tau_0}
\frac{\sin(\Omega_{\mathrm{eff},s}\tau_0)}{\Omega_{\mathrm{eff},s}}.
\end{equation}
The corresponding QFI is
\begin{equation}
\mathcal{F}_{Q,\mathrm{free}}
= 4|\partial_A\beta|^2
=
16\frac{\sin^2(\Omega_{\mathrm{eff},s}\tau_0)}{\Omega_{\mathrm{eff},s}^2}
=
16\tau_0^2\,\mathrm{sinc}^2(\Omega_{\mathrm{eff},s}\tau_0).
\end{equation}

\textbf{QFI for the geometric protocol.}
First, we would like to write the evolution operators in the same form as in Eq.~\eqref{UiaH}.
The displacement operator can be written as
\begin{equation}
D(\Sigma)
=
\exp(\Sigma \hat a^\dagger - \Sigma^* \hat a)
=
\exp[-iA\hat H_c],
\end{equation}
with
\begin{equation}
\hat H_c
=
\frac{i(\Sigma \hat a^\dagger - \Sigma^* \hat a)}{A}.
\end{equation}
Similarly, the qubit phase factor may be written as
\begin{equation}
e^{i\delta\Phi\,\sigma_z/2}
=
\exp[+iA\hat H_{\mathrm{qubit}}]
=
\exp[-iA(-\hat H_{\mathrm{qubit}})].
\end{equation}
As a result, the total evolution operator can be cast into the form
\begin{equation}
U(A)
=
\exp[-iA(\hat H_c-\hat H_{\mathrm{qubit}})].
\end{equation}
For a pure state, the QFI is four times the variance of the generator,
\begin{equation}
\mathcal{F}_Q
=
4\,\mathrm{Var}(\hat H_c-\hat H_{\mathrm{qubit}}).
\end{equation}
Using the identity
\begin{equation}
\mathrm{Var}(X-Y)=\mathrm{Var}(X)+\mathrm{Var}(Y)-2\,\mathrm{Cov}(X,Y),
\end{equation}
one can obtain
\begin{equation}
\begin{split}
    \mathcal{F}_{Q,\mathrm{geo}}
& =
4\left[
\mathrm{Var}(\hat H_c)
+
\mathrm{Var}(\hat H_{\mathrm{qubit}})
-
2\,\mathrm{Cov}(\hat H_c,\hat H_{\mathrm{qubit}})
\right] \\
& = 4\left[
\langle \hat H_c^2\rangle
-
\langle \hat H_c\rangle^2
+
\langle \hat H_{\mathrm{qubit}}^2\rangle
-
\langle \hat H_{\mathrm{qubit}}\rangle^2
-
2\left(
\langle \hat H_c \otimes \hat H_{\mathrm{qubit}}\rangle
-
\langle \hat H_c\rangle \langle \hat H_{\mathrm{qubit}}\rangle
\right)
\right].
\label{FQI_geo2}
\end{split}
\end{equation}
Although in general $4\,\mathrm{Var}(\hat H_c-\hat H_{\mathrm{qubit}})
\neq
4\,\mathrm{Var}(\hat H_c)+4\,\mathrm{Var}(\hat H_{\mathrm{qubit}})$,
they become equal when the covariance term vanishes.
More specifically, if the initial state is a direct product state of the cavity and qubit,
\begin{equation}
\rho_0=\rho_c\otimes \rho_q,
\end{equation}
then $\hat H_c$ and $\hat H_{\mathrm{qubit}}$ act on different Hilbert spaces, and
\begin{equation}
\langle \hat H_c \otimes \hat H_{\mathrm{qubit}}\rangle
=
\langle \hat H_c\rangle \langle \hat H_{\mathrm{qubit}}\rangle.
\end{equation}
The covariance terms in Eq.~\eqref{FQI_geo2} therefore vanish, and the QFI reduces to
\begin{equation}
\mathcal{F}_{Q,\mathrm{protocol}}
=
4\,\mathrm{Var}(\hat H_c)
+
4\,\mathrm{Var}(\hat H_{\mathrm{qubit}}).
\end{equation}

The qubit-sector generator is
\begin{equation}
\hat H_{\mathrm{qubit}}=\frac{\delta\Phi}{2A}\,\sigma_z.
\end{equation}
If the qubit is prepared in
\begin{equation}
\frac{1}{\sqrt{2}} (|e\rangle + |g\rangle),
\end{equation}
which is obtained from $\ket{g}$ by $R_y(-\pi/2)$ under the convention $\sigma_z\ket{e}=+\ket{e}$ and $\sigma_z\ket{g}=-\ket{g}$, one has
\begin{equation}
\langle \sigma_z \rangle = 0,
\qquad
\mathrm{Var}(\sigma_z)=1.
\end{equation}
Therefore, the qubit contribution to the QFI is
\begin{equation}
4\,\mathrm{Var}(\hat H_{\mathrm{qubit}})
=
4\cdot \frac{\delta\Phi^2}{4A^2}
=
\frac{\delta\Phi^2}{A^2}.
\end{equation}
With the total phase
\begin{equation}
\delta\Phi
=
A \alpha e^r \chi \tau_0^2 \sin(\Delta\tau_0-\phi_1)\,
\mathrm{sinc}\!\left(\frac{(\Delta+\chi/2)\tau_0}{2}\right)
\mathrm{sinc}\!\left(\frac{(\Delta-\chi/2)\tau_0}{2}\right),
\end{equation}
Averaging over the random relative phase, with $\langle\sin^2(\Delta\tau_0-\phi_1)\rangle=1/2$, yields Eq.~\eqref{FQI_geo}.

\appsectionnolabel{\texorpdfstring{The Impact of Decoherence from the Geometric Protocol}{The Impact of Decoherence from the Geometric Protocol}}
\label{app:background}
To rigorously evaluate the protocol's performance, we must account for environmental decoherence. The open-system dynamics of the joint qubit-cavity state $\rho(t)$ are governed by the Lindblad master equation:
\begin{equation}
    \dot{\rho}(t) = -i \left[ \frac{\chi}{2} \hat{a}^\dagger \hat{a} \hat{\sigma}_z, \rho(t) \right] + \kappa(\bar{n}_{\rm th}+1)\mathcal{D}[\hat{a}]\rho + \kappa \bar{n}_{\rm th}\mathcal{D}[\hat{a}^\dagger]\rho + \Gamma_1\mathcal{D}[\sigma_-]\rho + \frac{\Gamma_\phi}{2}\mathcal{D}[\sigma_z]\rho,
    \label{eq:full_master}
\end{equation}
where $\mathcal{D}[\hat{O}]\rho = \hat{O}\rho \hat{O}^\dagger - \frac{1}{2}\{\hat{O}^\dagger \hat{O}, \rho\}$. Here,  $\kappa\equiv 1/T_1^c$ is the cavity single-photon loss rate, $\bar{n}_{\rm th}$ is the thermal photon occupation, and $\Gamma_1$ and $\Gamma_\phi$ are the intrinsic qubit relaxation and pure dephasing rates, respectively.

We first evaluate the mean cavity amplitude $\langle\hat{a}\rangle=\text{Tr}(\hat{a}\rho)$ conditioned on $s=+1$ for $\ket{e}$ and $s=-1$ for $\ket{g}$, using the same convention as the main text. Using the bosonic commutation relation $[\hat{a}, \hat{a}^\dagger] = 1$, the coherent amplitude evolves as
\begin{equation}
    \frac{d}{dt}\langle\hat{a}\rangle= \frac{d}{dt}\text{Tr}(\hat{a}\dot\rho) = -\left(i \frac{s\chi}{2} + \frac{\kappa}{2}\right) \langle\hat{a}\rangle.
\end{equation}
The thermal photon number $\bar{n}_{\rm th}$ cancels exactly, indicating that the mean trajectory $\alpha_s(t)$ is independent of thermal noise and is damped only by the vacuum loss rate $\kappa/2$. Given the initial condition $\langle\hat{a}(0)\rangle = \alpha e^r \equiv \beta$, the cavity amplitude bifurcates into two spin-conditioned trajectories, $\alpha_+(t)$ and $\alpha_-(t)$, during the free evolution.

To capture thermal decoherence, we consider the initial density matrix $\rho_0 = |\psi\rangle\langle\psi| \otimes \hat{D}(\beta)\rho_{\rm th}\hat{D}^\dagger(\beta)$, where $\rho_{\rm th}$ is the thermal state and $\hat{D}(\beta)$ is the displacement operator. Because $\rho_{\rm th}$ is phase-invariant, the conditional cavity states evolve as $\rho_{\pm}(t) = \hat{D}(\alpha_\pm(t)) \rho_{\rm th} \hat{D}^\dagger(\alpha_\pm(t))$. Tracing over the cavity degrees of freedom evaluates the qubit off-diagonal element $\rho_{eg}$ at the end of the sequence:
\begin{equation}
    \text{Tr}_{\rm cav} \left( \hat{D}(\alpha_+(t)) \rho_{\rm th} \hat{D}^\dagger(\alpha_-(t)) \right) = e^{i\Phi(t)} \text{Tr}_{\rm cav} \left( \rho_{\rm th} \hat{D}(\Delta\alpha(t)) \right),
\end{equation}
where $\Delta\alpha(t) = \alpha_+(t) - \alpha_-(t) \simeq -2i\beta \sin(\chi t / 2)$ is the phase-space separation, and $e^{i\Phi(t)}$ is an irrelevant deterministic phase. Using the characteristic function of a thermal state, this factor becomes
\begin{equation}
    \text{Tr} \left( \rho_{\rm th} \hat{D}(\Delta\alpha) \right) = \exp\left[ -\frac{1}{2}(1 + 2\bar{n}_{\rm th})|\Delta\alpha(t)|^2 \right].
\end{equation}

To ensure valid geometric phase accumulation, the protocol duration must satisfy $2\tau_0 \ll T_1$ to avoid irreversible phase disruptions from spontaneous qubit relaxation. Integrating the instantaneous phase-space distinguishability over the full spin-echo sequence yields the effective decoherence envelope,
\begin{equation}
    \rho_{eg}(2\tau_0) = \rho_{eg}(0)e^{i\delta\phi} \exp\left[ -\frac{2\tau_0}{T_{2, \rm echo}^{(0)}} - \Lambda_\kappa(\tau_0, \bar{n}_{\rm th}) \right],
\end{equation}
where $T_{2, \rm echo}^{(0)}$ characterizes the intrinsic qubit decay during the spin-echo sequence. The cavity-induced dephasing exponent $\Lambda_\kappa$ integrates the squared phase-space separation weighted by the thermal bath factor,
\begin{equation}
    \Lambda_\kappa(\tau_0, \bar{n}_{\rm th}) = \frac{\kappa}{2} (1 + 2\bar{n}_{\rm th}) \int_0^{2\tau_0} |\Delta \alpha(t')|^2 dt'.
\end{equation}
Exploiting the symmetry of the echo sequence, we substitute $|\Delta \alpha(t')|^2 = 2|\beta|^2 [1 - \cos(\chi t')]$ to evaluate the integral analytically, yielding
\begin{equation}
    \Lambda_\kappa(\tau_0, \bar{n}_{\rm th}) = \kappa (1 + 2\bar{n}_{\rm th}) |\beta|^2 \tau_0 \left( 1 - \frac{\sin(\chi \tau_0)}{\chi \tau_0} \right).
\end{equation}

The decoherence of the final qubit signal can be decomposed into three physical mechanisms,
\begin{equation}
    \frac{1}{T_{2, \rm eff}} \simeq \underbrace{\frac{1}{T_{2, \rm echo}^{(0)}}}_{\text{Intrinsic Qubit Decay}} + \underbrace{\kappa |\beta|^2 \left( 1 - \frac{\sin(\chi \tau_0)}{\chi \tau_0} \right)}_{\text{Cavity Vacuum Dephasing}} + \underbrace{2\bar{n}_{\rm th} \kappa |\beta|^2 \left( 1 - \frac{\sin(\chi \tau_0)}{\chi \tau_0} \right)}_{\text{Cavity Thermal Dephasing}}.
\end{equation}

Each contribution to the effective decoherence rate represents a noise mechanism that limits the protocol performance.

The first term arises from the bare qubit decay. During the dynamic evolution, the central spin-echo sequence effectively filters out low-frequency fluctuations, leading to a baseline coherence time $T_{2, \rm echo}^{(0)}$ that is typically longer than the standard free induction decay time $T_2^*$.

The second term describes vacuum which-path dephasing caused by zero-temperature cavity loss, the dominant noise source in this protocol. Leaked photons carry information about the qubit branch, and the large intermediate photon occupation enhances the resulting dephasing.

The final term represents the thermal amplification driven by the finite thermal bath occupation. The residual thermal fluctuations interact with the large coherent state, amplifying the which-path dephasing by a multiplicative factor of $1+2\bar{n}_{\rm th}$ relative to the pure vacuum case. Operating a high-quality superconducting cavity at typical gigahertz frequencies within a dilution refrigerator environment ensures that the deep cryogenic regime $\bar{n}_{\rm th} \ll 1$ is readily achieved, thereby safely suppressing this thermal multiplier effect in practical experimental setups.

\appsectionnolabel{\texorpdfstring{Signal Profile and Experimental Parameter Choices}{Signal Profile and Experimental Parameter Choices}}
\label{singal profile}

In the main text, we choose $2\tau_0\approx\tau_{\rm DM}$ and $\chi\tau_0=\pi$ to match the DM linewidth and maximize the characteristic signal response.

The virialized DM speed follows the Maxwellian distribution of the Standard Halo Model,
\begin{equation}
	f_{\mathrm{DM}}(v)=\frac{v}{\sqrt{\pi}v_{vir} v_g}e^{-(v+v_g)^2/v_{vir}^2}(e^{4vv_g/v_{vir}^2}-1),
\end{equation}
which yields the DM frequency spectrum
\begin{equation}
	f_{\mathrm{DM}}(\omega_{\mathrm{DM}})=\frac{1}{ \sqrt{2({\omega_{\mathrm{DM}}}{m_{\mathrm{DM}}}-m_{\mathrm{DM}}^2)}}f_{\mathrm{DM}}\left[\sqrt{2\left(\frac{\omega_{\mathrm{DM}}}{m_{\mathrm{DM}}}-1\right)}\right].
\end{equation}
Here, $v_{vir}\approx 220$~km/s is the virial velocity and $v_g\approx 232$ km/s is the speed of the Sun relative to the halo rest frame. 

The DM profile is strictly zero for $\omega_{\mathrm{DM}}< m_{\mathrm{DM}}$, with a characteristic width $\sigma \sim 10^{-6} m_{\rm DM}$. The total signal power $P_{\rm DM}$ is obtained by convolving the monochromatic response from Eq.~\eqref{dPhi} with this spectral density:
\begin{equation}
P_{\rm DM} = \int  d\omega f_{\mathrm{DM}}(\omega) \frac{1}{2}(\alpha e^r \chi \tau_0^2)^2 \text{sinc}^2 \left( \frac{(\Delta + \chi/2)\tau_0}{2} \right) \text{sinc}^2 \left( \frac{(\Delta - \chi/2)\tau_0}{2} \right).
\label{eq:geoscaling}
\end{equation}

In the short-time limit ($\tau_0 \ll \sigma^{-1}, \chi^{-1}$), the $\text{sinc}$ functions approach unity, and the signal power scales as $P_{\rm DM} \propto \tau_0^4$. In this regime, the probe's energy resolution is too broad to resolve the DM spectral features. Conversely, in the long-time limit ($\tau_0 \gg \sigma^{-1}, \chi^{-1}$), the response is dominated by the resonance poles at $\Delta = \pm \chi/2$ , where $\Delta=\omega_c-\omega_{\rm DM}$. As $\tau_0 \to \infty$, the function $\dfrac{\sin^2(x \tau_0/2)}{(x/2)^2}$ converges to the Dirac delta distribution $2\pi \tau_0 \delta(x)$. Rewriting the signal power to isolate these poles, we have:
\begin{equation}
P_{\rm DM} = \int d\Delta f_{\rm DM}(\omega_c-\Delta) \frac{8 \alpha^2 e^{2r} \chi^2}{(\Delta^2 - \chi^2/4)^2} \sin^2\left(\frac{(\Delta + \chi/2)\tau_0}{2}\right) \sin^2\left(\frac{(\Delta - \chi/2)\tau_0}{2}\right).
\end{equation}
Evaluating the integral around $\Delta=\pm\chi/2$ gives the asymptotic long-time behavior. Because $f_{\rm DM}$ is defined as a function of the absolute DM frequency, the two spectral factors are $f_{\rm DM}(\omega_c\mp\chi/2)$,
\begin{equation}
P_{\rm DM} \approx 4\pi \alpha^2 e^{2r} \tau_0 \sin^2\left(\frac{\chi \tau_0}{2}\right) \Big[ f_{\rm DM}(\omega_c+\chi/2)\Theta(\omega_c+\chi/2-m_{\rm DM}) + f_{\rm DM}(\omega_c-\chi/2) \Theta(\omega_c-\chi/2-m_{\rm DM})\Big].
\end{equation}
This linear scaling with $\tau_0$ reflects incoherent power addition once the protocol duration exceeds the DM coherence time. To optimize the signal, we consider a Gaussian approximation for the DM line shape, $f_{\rm DM}(\Delta) \propto \exp[-(\Delta - \Delta_0)^2 / 2\sigma^2]$, where $\Delta_0 = \omega_c - \omega_0$ is the cavity detuning from the DM peak. The spectral sampling term $S = f_{\rm DM}(+\chi/2) + f_{\rm DM}(-\chi/2)$ becomes:
\begin{equation}
S \propto \exp\left[-\frac{(\chi/2 - \Delta_0)^2}{2\sigma^2}\right] + \exp\left[-\frac{(\chi/2 + \Delta_0)^2}{2\sigma^2}\right].
\end{equation}
Maximizing $S$ with respect to $\Delta_0$ gives $\Delta_0=0$, so the response is largest when the cavity frequency is aligned with the peak of the DM spectrum. Under this condition, $S \propto \exp(-\chi^2 / 8\sigma^2)$, which imposes a constraint on the dispersive coupling: $\chi$ must be comparable to or smaller than the DM linewidth ($\chi \lesssim \sigma\approx 1/\tau_{\rm DM}$) to prevent the resonance poles from being pushed into the spectral tails. Finally, to maximize the $\sin^2(\chi \tau_0/2)$ factor, the protocol duration is ideally set to $\tau_0 = \pi / \chi$.

\begin{figure}[htp!]
    \centering
    \includegraphics[width=0.6\linewidth]{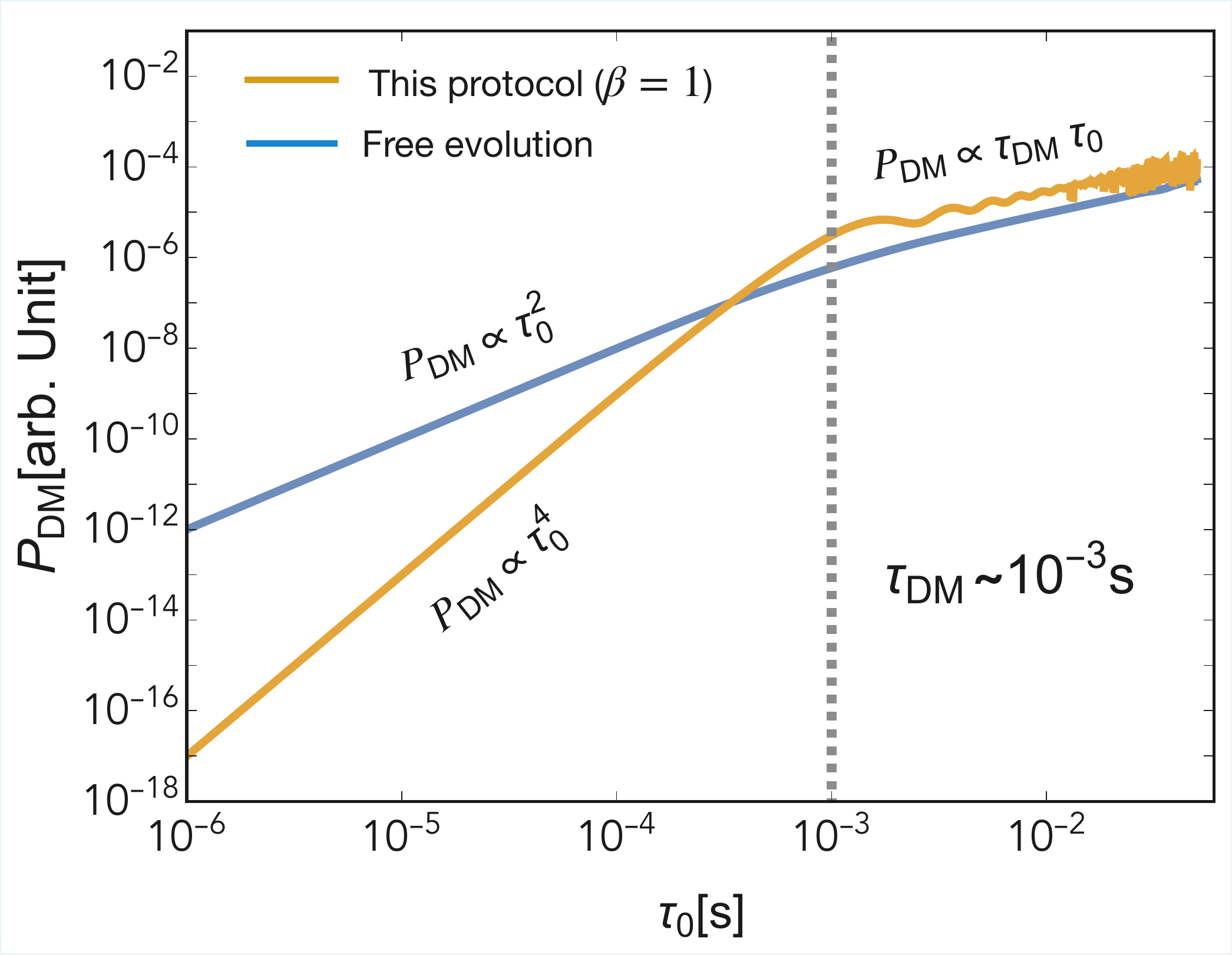}
    \caption{Signal power as a function of the protocol time for the benchmark parameters specified in the main text. Both protocols exhibit distinct scaling below and above the DM coherence time.}
    \label{fig:scaling}
\end{figure}

Figure~\ref{fig:scaling} compares the dependence of the geometric-protocol signal power in Eq.~\eqref{eq:geoscaling} on $\tau_0$ with that of free evolution. We use a benchmark DM frequency of $1\,\mathrm{GHz}$, the parameters specified above, and $\beta=1$. The free-evolution signal power is

\begin{equation}
    P_{\rm DM,s}^{\rm free}(\tau) = |A|^2\int d\omega\,f_{\rm DM}(\omega) \left(\frac{\sin(\Omega_{\rm eff,s}\tau/2)}{\Omega_{\rm eff,s}/2}\right)^2.
\end{equation}

For free evolution, the signal power scales as $\tau^2$ for $\tau\ll\tau_{\rm DM}$ and as $\tau\tau_{\rm DM}$ for $\tau\gg\tau_{\rm DM}$, where $\tau_{\rm DM}\sim10^{-3}\,\mathrm{s}$ at the benchmark frequency. The geometric protocol scales as $\tau_0^4$ below the coherence time because of its double-sinc response and approaches the free-evolution scaling above $\tau_{\rm DM}$. This crossover supports the choice $2\tau_0\approx\tau_{\rm DM}$.

\appsectionnolabel{Power Spectrum Density}
\label{app:PSD}

In this appendix, we provide a detailed derivation of the power spectral density (PSD) and the corresponding quantum noise for a single qubit sensor. 

The protocol uses a single qubit for readout. Let $M_j\equiv\hat\sigma_j^y/2$ denote the outcome operator for the $j$th measurement. Its expectation value is $\langle M_j\rangle=\mathrm{Tr}(\rho_jM_j)=F_j$, where $\rho_j$ is the qubit state at time step $j$. The signal $F_j$ follows from Eqs.~\eqref{dPhi} and \eqref{soperator}, including the dephasing factor and the phase update $\phi_1\to\phi_1+\Delta j\tau$. In the absence of a signal, $F_j=0$. The two-point correlation function is $C_{jj'}=\mathrm{Tr}(\rho_{jj'}M_jM_{j'})$, with $\rho_{jj'}\equiv\rho_j\otimes\rho_{j'}$. In the zero-signal limit, independent measurements give
$$
C_{jj'} \simeq \frac{1}{4}\delta_{jj'}
$$
where the factor $1/4$ originates from the variance of an unpolarized single qubit state. 

The discrete power spectral density (PSD) operator is defined via the Fourier transform of the time-sequence measurement outcomes as
$$
\mathcal{O}_k \equiv \frac{\tau^2}{t_{\mathrm{obs}}}\sum_{j,j'} e^{2\pi i k (j-j')/N_{\mathrm{obs}}} M_j M_{j'}
$$
where $\tau$ is the free precession time per measurement, which in this protocol is taken to be $\tau=2\tau_0$. $N_{\mathrm{obs}}$ is the total number of measurements, $t_{\mathrm{obs}} = N_{\mathrm{obs}}\tau$ is the total observation time, and $\omega_k = 2\pi k / t_{\mathrm{obs}}$. The expectation value of the PSD, $\mathcal{P}_k \equiv \langle \mathcal{O}_k \rangle$, in the zero-signal background evaluates to
$$
\langle \mathcal{O}_k \rangle|_{A=0} = \frac{\tau^2}{t_{\mathrm{obs}}}\sum_{j,j'} e^{2\pi i k (j-j')/N_{\mathrm{obs}}} \frac{1}{4}\delta_{jj'} = \frac{\tau^2}{N_{\mathrm{obs}}\tau} \frac{N_{\mathrm{obs}}}{4} = \frac{\tau}{4}
$$
This demonstrates that the mean projection noise background is $\tau/4$ and is strictly independent of the frequency bin $k$. Consequently, the true signal PSD is defined by subtracting this constant uniform background, $\mathcal{S}_k \equiv \mathcal{P}_k - \tau/4$.

The background PSD fluctuation that enters the test statistic calculation is given by $\mathcal{B}_k \equiv \sqrt{\langle \mathcal{O}_k^2 \rangle - \langle \mathcal{O}_k \rangle^2}|_{A=0}$. Evaluating $\langle \mathcal{O}_k^2 \rangle$ requires calculating the four-point correlation function of the measurement operator $M_j^z$. Expanding the square of the PSD operator yields
$$
\langle \mathcal{O}_k^2 \rangle = \frac{\tau^4}{16 t_{\mathrm{obs}}^2} \left\langle \left( \sum_{j} \mathbf{1} + \sum_{j \neq j'} e^{2\pi i k (j-j')/N_{\mathrm{obs}}} \hat\sigma_j^y \hat\sigma_{j'}^y \right)^2 \right\rangle
$$
Because the trace of an odd number of Pauli matrices over the unpolarized state vanishes, the non-zero contributions only arise from terms proportional to the identity operator. Expanding the parenthesis, the first term trivially yields $N_{\mathrm{obs}}^2$. The cross-term squared gives rise to a summation $\sum_{j_1 \neq j_2} \sum_{j_3 \neq j_4} e^{2\pi i k (j_1-j_2+j_3-j_4)/N_{\mathrm{obs}}} \hat\sigma_{j_1}^y \hat\sigma_{j_2}^y \hat\sigma_{j_3}^y \hat\sigma_{j_4}^y$. For the trace to survive, the indices must pair up perfectly (i.e., $j_1=j_3, j_2=j_4$ or $j_1=j_4, j_2=j_3$). This pairing collapses the four-fold summation into
$$
\sum_{j_1 \neq j_2} \left( 1 + e^{4\pi i k (j_1-j_2)/N_{\mathrm{obs}}} \right) = N_{\mathrm{obs}}(N_{\mathrm{obs}}-1) + N_{\mathrm{obs}}(N_{\mathrm{obs}}\delta_{k,0} - 1)
$$
where the Kronecker delta $\delta_{k,0}$ isolates the zero-frequency mode. Summing these non-vanishing contributions and neglecting subleading terms of $\mathcal{O}(N_{\mathrm{obs}})$, the evaluation leads to a distinct separation between the zero-frequency mode and all other finite frequency modes. The resulting standard deviations are derived as
\begin{align}
\mathcal{B}_{k} \simeq \frac{\tau}{4} \quad\left(\frac{\tau}{2\sqrt{2}}~{\rm for}~k=0\right).  
\end{align}
These derived results align with the relation of mean and standard deviation value of  $\chi^2$ distribution with 2 degrees (1 degree for $k=0$) freedom.

The diagonal contribution to $\mathcal P_k$ equals the background level $\mathcal B_k$ for $k\ne0$ and is removed by the background subtraction. We therefore retain only the off-diagonal signal contribution.

The expected signal in the $j$th measurement is proportional to $\cos(\Delta\tau_0+\phi_{1j})$, where $\phi_{1j}=\phi_1+\Delta j\tau$ and $\phi_1$ is the random DM phase at the beginning of the measurement sequence. Because the field has a finite coherence time $\tau_{\rm DM}$, its phase remains correlated only for $|t-t'|\leq\tau_{\rm DM}$. For the choice $\tau\sim\tau_{\rm DM}$, only neighboring measurements have a nonzero signal correlation. Averaging over the random phases gives
\begin{align}
C_{\mathrm{sig}}(t,t') &=\dfrac{1}{2\pi}\int d\phi_1\dfrac{1}{2\pi}\int d\phi_1' \mathrm{Tr}[\rho_{jj'} M_j M_{j'}] \times[\Theta(|t_j-t_{j'}|-\tau_{\rm DM})+2\pi \delta(\phi_{1}-\phi_1')\Theta(\tau_{\rm DM}-|t_j-t_{j'}|)] \nonumber\\
 &=\mathcal{A}\cos\!\big[\Delta(t-t')\big]\Theta(\tau_{\rm DM}-|t-t'|),
\end{align}
where the coefficient of the correlated signal component is
\begin{align}
    \mathcal{A}=\frac{1}{8}A^2 \beta^2 \chi^2 \tau_0^4\sinc^2\left(\frac{(\Delta+\chi/2)\tau_0}{2}\right) \sinc^2\left(\frac{(\Delta-\chi/2)\tau_0}{2}\right)\eta^2.
\end{align}
Here $\eta$ is the decoherence factor defined in Eq.~\eqref{eta}. With our benchmark parameter choices, this expression simplifies to
\begin{equation}
\mathcal{A}
=
\frac{8}{\pi^2}
\beta^2 \tau_0^{2}
A^{2}\eta^2,
\end{equation}
where we used Eq.~\eqref{couplingA} and Eq.~\eqref{couplinga}, and evaluated the $\sinc$ factors in Eq.~\eqref{dPhi} at $\chi \tau_0 = \pi$ and $\Delta = 0$, so that $\sinc(\pi/4) = 2\sqrt{2}/\pi$.

Substituting this into the continuum expression for the PSD, we obtain
\begin{equation}
S_k =
\frac{1}{t_{\mathrm{obs}}}\int_0^{t_{\mathrm{obs}}} dt
\int_0^{t_{\mathrm{obs}}} dt'\,
e^{i\omega_k (t-t')} \mathcal{A}\cos\!\big[\Delta (t-t')\big]\Theta(\tau_{\rm DM}-|t-t'|).
\end{equation}

To simplify the double integral, it is convenient to introduce the relative and average time coordinates $u=t-t', v=\frac{t+t'}{2}$.
Since the integrand depends only on the relative time $u$, the integral over $v$ can be carried out immediately, which yields the standard identity
\begin{equation}
\int_0^{t_{\mathrm{obs}}} dt \int_0^{t_{\mathrm{obs}}} dt'\, f(t-t')
=
\int_{-t_{\mathrm{obs}}}^{t_{\mathrm{obs}}} du\, (t_{\mathrm{obs}}-|u|)\, f(u).
\end{equation}
Therefore,
\begin{equation}
S_k=
\frac{\mathcal{A}}{t_{\mathrm{obs}}}
\int_{-t_{\mathrm{obs}}}^{t_{\mathrm{obs}}} du\,
(t_{\mathrm{obs}}-|u|)\,
e^{i\omega_k u}\cos(\Delta u)\,
\Theta(\tau_{\rm DM}-|u|).
\end{equation}

At this point, the physical structure of the result becomes transparent. The factor
$e^{i\omega_k u}\cos(\Delta u)$ mixes the Fourier-bin frequency $\omega_k$ with the signal frequency $\Delta$, while the step function enforces the finite coherence time. We now rewrite the cosine in exponential form,
\begin{equation}
e^{i\omega_k u}\cos(\Delta u)
=
\frac{1}{2}\left(e^{i(\omega_k-\Delta)u}+e^{i(\omega_k+\Delta)u}\right).
\end{equation}
Defining
\begin{equation}
\Delta\omega_k \equiv \omega_k - \Delta,
\end{equation}
the term proportional to $e^{i(\omega_k+\Delta)u}$ oscillates rapidly and averages to zero. Retaining the near-resonant contribution gives the usual narrow-band approximation. The factor $1/2$ from the exponential representation of the cosine remains explicit, so
\begin{equation}
S_k \simeq
\frac{\mathcal{A}}{2t_{\mathrm{obs}}}
\int du\,
(t_{\mathrm{obs}}-|u|)\cos(\Delta\omega_k u)\,
\Theta(\tau_{\rm DM}-|u|).
\end{equation}

It is convenient to denote the effective integration range by $\Lambda \equiv \min(t_{\mathrm{obs}},\tau_{\rm DM})$. It indicates that, if the single measurement time is shorter than the DM coherence time, the DM can maintain a coherent phase during the measurement and one can do the integration over the whole measurement time. Otherwise, it is necessary to separate the whole measurement time into different patches of DM coherence time. With this integration limit, the signal spectrum is,
\begin{equation}
S_k \simeq
\frac{\mathcal{A}}{2t_{\mathrm{obs}}}
\int_{-\Lambda}^{\Lambda} du\,
(t_{\mathrm{obs}}-|u|)\cos(\Delta\omega_k u).
\end{equation}
From here, the derivation naturally splits into the following two cases.

\subsection{Case I: $t_{\mathrm{obs}}<\tau_{\rm DM}$}

If the total observation time is shorter than the coherence time, the signal remains phase-coherent throughout the entire measurement window. In such a case,
\begin{equation}
\Theta(\tau_{\rm DM}-|u|)=1,
\qquad |u|\le t_{\mathrm{obs}},
\end{equation}
Because the integrand is even in $u$, we can write the PSD as
\begin{equation}
S_k=
\frac{\mathcal{A}}{t_{\mathrm{obs}}}
\int_0^{t_{\mathrm{obs}}} du\,
(t_{\mathrm{obs}}-u)\cos(\Delta\omega_k u).
\end{equation}

To evaluate the remaining integral, we define
\begin{equation}
I(T,\Delta)\equiv \int_0^T (T-u)\cos(\Delta u)\,du = T\int_0^T \cos(\Delta u)\,du - \int_0^T u\cos(\Delta u)\,du.
\end{equation}
The two integrals are
\begin{equation}
\int_0^T \cos(\Delta u)\,du = \frac{\sin(\Delta T)}{\Delta} \,, \quad \int_0^T u\cos(\Delta u)\,du
=
\frac{T\sin(\Delta T)}{\Delta}
+
\frac{\cos(\Delta T)-1}{\Delta^2}
.
\end{equation}
Substituting these results back, we find
\begin{equation}
I
=
T\frac{\sin(\Delta T)}{\Delta}
-
\left[
\frac{T\sin(\Delta T)}{\Delta}
+
\frac{\cos(\Delta T)-1}{\Delta^2}
\right]
=
\frac{1-\cos(\Delta T)}{\Delta^2}
=\frac{2\sin^2(\Delta T/2)}{\Delta^2}.
\end{equation}
In the last equality, we used $1-\cos x=2\sin^2(x/2)$.
Setting $T=t_{\mathrm{obs}}$ and $\Delta=\Delta\omega_k$, the PSD becomes
\begin{equation}
S_k \simeq
\frac{2\mathcal{A}}{t_{\mathrm{obs}}\Delta\omega_k^2}
\sin^2\!\left(\frac{t_{\mathrm{obs}}\Delta\omega_k}{2}\right).
\end{equation}

\subsection{Case II: $t_{\mathrm{obs}}>\tau_{\rm DM}$}
In the experimentally relevant regime $t_{\mathrm{obs}}>\tau_{\rm DM}$, the signal loses phase memory for separations larger than $\tau_{\rm DM}$. Only the interval $|u|<\tau_{\rm DM}$ contributes,
\begin{equation}
S_k=
\frac{\mathcal{A}}{t_{\mathrm{obs}}}
\int_0^{\tau_{\rm DM}} du\,
(t_{\mathrm{obs}}-u)\cos(\Delta\omega_k u).
\end{equation}
Here again we use the evenness of the integrand,
The two parts of the integration follow the same structure as mentioned in the above case, here we directly write the result as,
\begin{equation}
S_k=
\frac{\mathcal{A}}{t_{\mathrm{obs}}}
\left[
\frac{(t_{\mathrm{obs}}-\tau_{\rm DM})\sin(\Delta\omega_k\tau_{\rm DM})}{\Delta\omega_k}
+
\frac{1-\cos(\Delta\omega_k\tau_{\rm DM})}{\Delta\omega_k^2}
\right].
\end{equation}
Using again $1-\cos x = 2\sin^2\!\left(\frac{x}{2}\right)$,
we get
\begin{equation}
S_k \simeq
\frac{2\mathcal{A}}{t_{\mathrm{obs}}\Delta\omega_k^2}
\sin^2\!\left(\frac{\tau_{\rm DM}\Delta\omega_k}{2}\right)
+
\frac{t_{\mathrm{obs}}-\tau_{\rm DM}}{t_{\mathrm{obs}}\Delta\omega_k}
\mathcal{A}\sin(\tau_{\rm DM}\Delta\omega_k).
\end{equation}
This expression agrees with Refs.~\cite{Chigusa:2024psk,Ma:2025orp} and is used in the main text to compute the projected 95\% C.L. sensitivity.

\end{document}